\documentclass[usenatbib]{mn2e}  
\usepackage{natbib}
\usepackage{graphicx}
\usepackage{amsmath}
\usepackage{amssymb}
\usepackage{txfonts}
\usepackage{multirow}
\def\hii{\mbox{H\,{\sc ii}}}

   \title[SNR candidate G351.7--1.2]{Non-thermal emission from massive star forming regions: A possible SNR candidate G351.7--1.2?}
 
\author[Veena et al.]{V. S. Veena$^{1}$\thanks{E-mail: veenavs.13@iist.ac.in}, S. Vig$^1$, B. Sebastian$^2$, D. V. Lal$^2$, A. Tej$^1$, S. K. Ghosh$^3$\\
$^1$Indian Institute of Space science and Technology, Thiruvananthapuram, 695 547, India\\
$^2$National Centre for Radio Astrophysics (NCRA-TIFR), Pune, 411 007, India\\
$^3$Tata Institute of Fundamental Research, Mumbai, 400 005, India}

\begin{document}

\date{}

\pagerange{\pageref{firstpage}--\pageref{lastpage}} \pubyear{}

\maketitle

\label{firstpage}

\begin{abstract}
We present low frequency wide band observations ($300-500$~MHz) of the star forming complex G351.7--1.2 using upgraded Giant Metrewave Radio Telescope (uGMRT), India. Combining this with the optical, infrared and submillimeter data, we analyse the large scale diffuse radio emission associated with the region that exhibits a broken shell morphology. The spectral index of the emission in the shell is $-0.8$, indicating non-thermal emission. H$\alpha$ emission that mimics the morphology of the radio shell on a smaller scale is also detected here. Based on the non-thermal emission from the radio shell and the presence of its optical counterpart, we classify G351.7--1.2 as a candidate SNR. A $\gamma$-ray source detected by $Fermi$ LAT (1FGLJ1729.1--3641c) is located towards the south-west of the radio shell and could have a possible origin in the interaction between high velocity particles from the SNR and the ambient molecular cloud. 
\end{abstract}


\section{Introduction}

The formation and evolution of massive OB stars (M$\geq$8~M$_\odot$) play a pivotal role in the evolution of the interstellar medium (ISM) and the host galaxies. This is primarily due to the enhanced feedback mechanisms from these stars that alter the environments  on local, global and cosmic scales. \hii~regions are one of the important classes of astrophysical objects that serve as a lodestar to identify regions of recent massive star formation activity. Formed around young OB stars that are still embedded in their natal cloud, these photoionized nebulae emit significantly in the longer wavelength bands, that are able to penetrate thick layers of gas and dust. In the earlier evolutionary phases, they are relatively small (size$<$0.1~pc) and dense (n$_e\geq10^4$~cm$^{-3}$) and are known as hypercompact and ultracompact \hii~regions. As they evolve, they transform into compact and classical \hii~regions \citep{2007prpl.conf..181H}.

\par Some of the Galactic ultracompact \hii~regions are found to possess large scale diffuse envelopes \citep[e.g.,][]{{1973MNRAS.162P...5H},{1996JKASS..29..177K},{1999ApJ...514..899H}}. The low density radio envelopes are likely to be associated with the observed \hii~regions. On the contrary, it is also possible that the emission is completely  unrelated to these objects. \citet{1999ApJ...514..232K} carried out a radio continuum survey to identify extended envelopes around a sample of ultracompact \hii~regions.  Of their 15 targets, 12 are found to possess extended envelopes. Based on the source morphologies, they concluded that atleast half of the extended components are directly connected with the ultracompact components. Using radio continuum observations, \citet{2001ApJ...549..979K} identified extended envelopes around 16 ultracompact \hii~regions. The velocity data obtained using radio recombination line observations suggested the physical association of these extended envelopes to the star forming regions. The preliminary step towards understanding the origin of the diffuse envelopes is to determine the properties of the large scale emission such as the emitting frequency and the spectral index. Such a study will provide vital clues on various physical processes and complex interaction mechanisms prevailing in these regions. Despite their significance, there is clearly a lack of detailed investigations on extended envelopes of ultracompact and compact \hii~regions  in the radio regime. In view of this, we have carried out a low frequency mapping study of a star forming complex, G351.7-1.2, in the southern sky. 

\begin{figure*}
\centering
\includegraphics[scale=0.55]{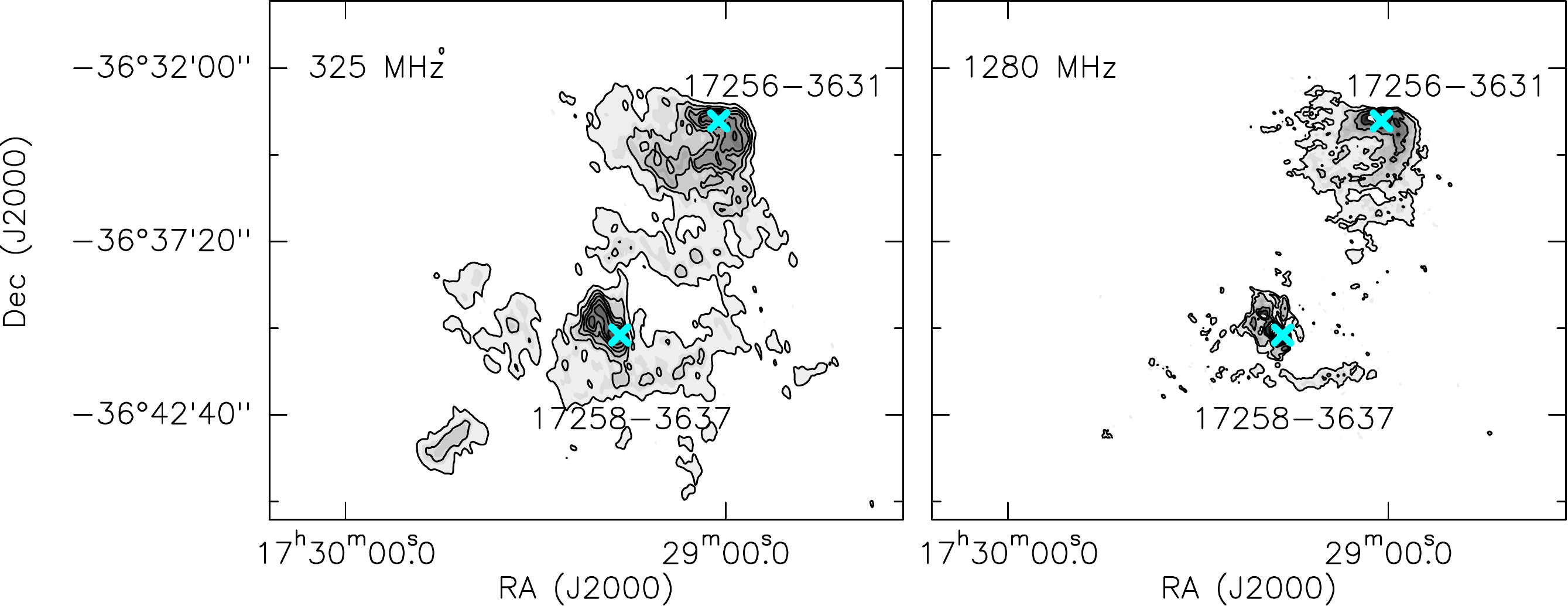}
\caption{(Left) 325~MHz radio continuum map of G351.7--1.2. The contour levels are from 22~mJy/beam to 250~mJy/beam in steps of 30~mJy/beam. The beam size is $21.7\arcsec\times8.5\arcsec$. (Right) 1280 MHz radio image of G351.7--1.2. The contour levels are 3.5, 8, 18, 30, 60, 120 and 180~mJy/beam. The beam size is $6.7\arcsec\times4.5\arcsec$ \citep{2017MNRAS.465.4219V}. The peak positions of two \hii~regions IRAS 17256--3631 and IRAS 17258--3637 are marked as $\times$.}
\label{gmrtold}
\end{figure*} 

\par The massive star forming complex G351.7--1.2 includes two \hii~regions: IRAS 17258--3637 and IRAS 17256--3631 (Fig.~\ref{gmrtold}). The former is a bipolar \hii~region. The star formation activity and gas kinematics towards this region have been investigated in detail earlier \citep{{2014MNRAS.440.3078V},{2017MNRAS.465.4219V}}. The latter is a \hii~region displaying cometary morphology. Our group has also extensively probed the structure, evolution and cometary morphology of this region \citep[][]{{2016MNRAS.456.2425V},{2017MNRAS.465.4219V}}. Both these regions are active sites of high mass star formation that harbour young embedded clusters. The low frequency radio continuum map of this region at 325~MHz revealed the presence of large scale diffuse emission connecting IRAS 17256--3631 and IRAS 17258--3637 \citep[Fig.~\ref{gmrtold} (left), taken from][]{2017MNRAS.465.4219V}. In order to examine the nature of this large scale emission, we probed this region again with improved sensitivity and larger bandwidth at low radio frequencies.

\par The wideband radio observations of the star forming complex associated with G351.7--1.2 carried out with the upgraded facilities of Giant Metrewave Radio Telescope (uGMRT) are presented in this work. We use the $300-500$~MHz band to map the large scale diffuse emission as well as to study the variation of spectral indices across the star forming regions. Combining this with the optical, infrared and submillimeter data, we present a multiwavelength investigation of the region that spans $14'$ in size. The organisation of the paper is as follows. The details of radio continuum observations and other archival data used in the study are given in Section 2. The results of our multiwavelength analysis are presented in Section 3 and summarized in Section 4.


\section{Observations and data reduction}
\subsection{Radio continuum observations using upgraded GMRT}

The ionized gas emission from the region associated with G351.7--1.2 is mapped using GMRT, India \citep{1991CuSc...60...95S}. GMRT consists of 30 antennas each having a diameter of 45~m arranged in a Y-shaped configuration. Twelve antennas are distributed randomly in a central array within an area of 1 square kilometre and the remaining antennas are stretched out along three arms each of length 14 km. The shortest and longest baselines are 105 m and 25~km respectively. This allows the mapping of high-resolution small, and low-resolution diffuse, large scale structures simultaneously with reasonable sensitivity. The bandwidth of observations with the upgraded GMRT \citep{gupta2017upgraded} was 200 MHz, between 300--500~MHz. The angular scale of the largest structure observable with the upgraded GMRT at 300 MHz is $\sim$35~$\arcmin$. The radio sources 3C286 and 3C48 were used as the primary flux calibrators as well as
the bandpass calibrators. The phase calibration source, 1830--360 was observed typically once every 35 minutes in order to correct for variations in the amplitude and phase across time.

\par We carried out the data reduction using the NRAO's Astronomical Image Processing System ($\tt{AIPS}$).
The spectral channels affected due to radio frequency interference, and the visibilities corresponding to non-working antennas were identified and flagged. The calibrated target data were split into six sub-bands each of bandwidth $\sim$32 MHz. The resulting central frequencies of each of these sub-bands are 321, 351, 385, 418, 450 and 480 MHz. All these six sub-bands were deconvolved and imaged separately using the task IMAGR and subjected to several iterations of self-calibration with a final round of amplitude and phase self-calibration to minimize the amplitude and phase errors.  Thus, we constructed six radio images at the above mentioned six central frequencies by limiting the ($u,v$)-range between 0.1 and 21 k$\lambda$ with identical synthesised beams of 19.9$\arcsec$ $\times$ 10.5$\arcsec$, the largest beam among the six frequencies considered. System temperature corrections were applied to account for the Galactic plane emission at these low radio frequencies. This is to account for the calibration of flux densities using a calibrator that is located away from the Galactic plane and the emission from the latter, which is significant at low radio frequencies. The factor $\rm{(T_{gal}+T_{sys})/T_{sys}}$ has been used to scale the flux density at each frequency, where T$_{\rm sys}$ is the system temperature corresponding to the flux density calibrator and T$_{\rm gal}$ is the temperature corresponding to the emission from the Galactic plane at the given frequency. We have used the radio continuum brightness of the whole sky at a frequency of 408 MHz \citep{1982A&AS...47....1H} and assumed a spectral index of $-$2.6 \citep{{1999A&AS..137....7R},{2011A&A...525A.138G}} to estimate T$_{\rm gal}$ at each of the six radio frequencies. The T$_{\rm sys}$ values are obtained by interpolating the system temperature values of upgraded GMRT \citep{gupta2017upgraded}. These measurements, T$_{\rm gal}$ and T$_{\rm sys}$ at each frequency were used to estimate the scale factor, which is then applied to the corresponding image at that frequency. Finally, these flux-density scaled images were corrected for the primary beam using
the task PBCOR. These `final' images, corrected for both, emission due to the Galactic plane and primary beam shapes have rms noises of 1.0, 0.9, 0.8, 0.8, 0.7 and and 0.7 mJy~beam$^{-1}$ at 321, 351, 385, 481, 450 and 480 MHz, respectively.

\begin{figure*}
\centering
\hspace*{-0.5cm}
\includegraphics[scale=0.57]{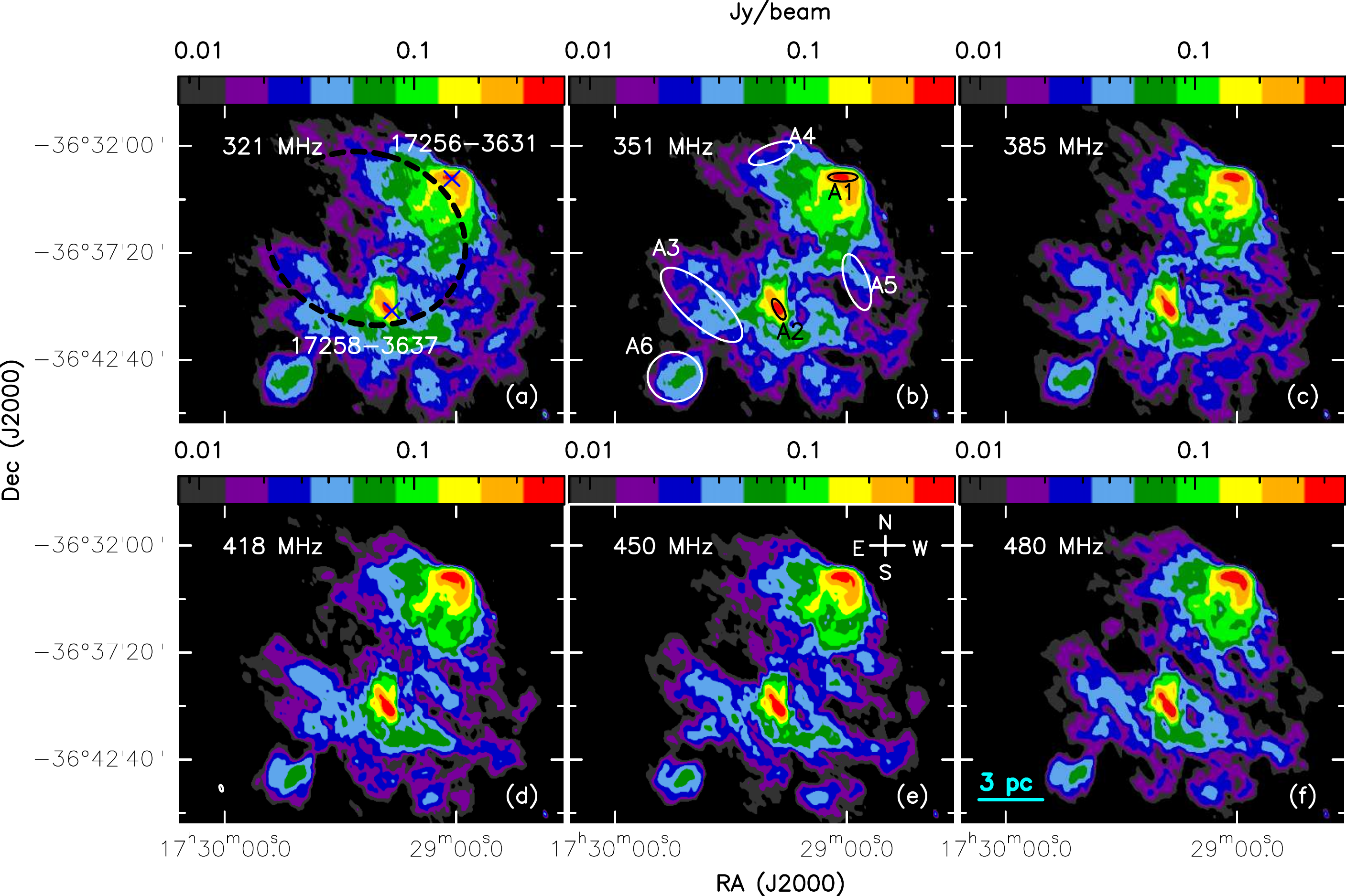}
\caption{Radio continuum maps of the star forming complex G351.7--1.2 at 321, 351, 385, 418, 450 and 480~MHz.  The two \hii~regions associated with this complex, IRAS 17256--3631 and IRAS 17258--3637 are marked and labelled in panel (a). The dashed arc shows the approximate position of diffuse ionised shell. Panel (b) shows 6 apertures whose spectral indices are listed in Table~\ref{snrtb}. The beam is shown as an ellipse in the bottom left corner of panel (d).}
\label{ugmrt}
\end{figure*} 

\subsection{Archival data}

The radio data from GMRT is complemented with data from various surveys. The region is probed in the optical wavebands using H$\alpha$+[NII] image obtained from the SuperCOSMOS H-alpha survey \citep{2005MNRAS.362..689P}. In order to investigate the emission at mid-infrared bands, we used images from the Midcourse Space Experiment (MSX) \citep{1999ASPC..177..394P} that mapped the Galactic plane at 4.3, 8.28, 12.13, 14.65, and 21.3~$\mu$m. We also used 4.5~$\mu$m image of this region observed by \textit{Spitzer Space Telescope} which is a Level-2 Post-Basic Calibrated data (PBCD) image from the Galactic Legacy Infrared Mid-Plane Survey \citep[GLIMPSE;][]{2003PASP..115..953B}. In addition, images from the \textit{Herschel Space Observatory} are used to study the emission from cold dust. The images are part of the \textit{Herschel} Hi-Gal Survey \citep{2010PASP..122..314M}. The instruments used in the survey are the Spectral and Photometric Imaging Receiver \citep[SPIRE;][]{2010A&A...518L...3G} and Photodetector Array Camera and Spectrometer \citep[PACS,][]{2010A&A...518L...2P}. The Hi-Gal observations were carried out in parallel mode covering wavelengths $70 - 500~\mu$m. We used Level-3 SPIRE images at 250, 350 and 500~$\mu$m for our analysis as the entire region was not covered by PACS at 70 and 160~$\mu$m. The resolution at 250, 350 and 500~$\mu$m are 18.1$\arcsec$, 24.9$\arcsec$ and 35.6$\arcsec$ respectively. We have also used the Apex+Planck image which is a combination of 870~$\mu$m data from the ATLASGAL survey \citep{2009A&A...504..415S} and 850~$\mu$m map from the Planck/HFI instrument \citep{2016A&A...585A.104C}. The data covers emission at large angular scales ($\sim$0.5$^\circ$), thereby revealing the structure of cold Galactic dust in greater detail. The pixel size and resolution achieved are 3.4$\arcsec$ and 21$\arcsec$, respectively. 

\section{Results and Discussions}
\subsection{Ionised radio shell}

The radio continuum maps of G351.7--1.2 at the six frequencies are shown in Fig.~\ref{ugmrt}. The \hii~regions IRAS 17256--3631 and IRAS 17258--3637 are marked and labelled in the images. Apart from the \hii~regions, the low frequency radio maps also reveal a large broken shell-like structure, highlighted with a dashed curve in the figure. This radio shell has not been identified in any of the previous Galactic plane surveys. The shell is $\sim14\arcmin$ in diameter, and has an opening towards the north-east. The physical size of the shell is 8.1~pc considering the distance to be $\sim$2~kpc. The latter is based on the  assumption that the shell is located at the same distance as the \hii~regions \citep[2~kpc,][]{{2014MNRAS.440.3078V},{2016MNRAS.456.2425V}}. At this distance, the shell could be $\sim43$~pc below the Galactic plane. The emission from the radio shell is highly clumpy and fragmented in nature. A striking feature is that the emission from the shell is more prominent at lower frequencies ($\nu\leq385$~MHz) unlike the emission from \hii~regions that become brighter as we move towards the high frequency bands. In order to closely examine the nature of emission from the newly discovered radio shell, we carried out a spectral index analysis. This is described in the next sub-section.

\begin{figure*}
\centering
\includegraphics[scale=0.27]{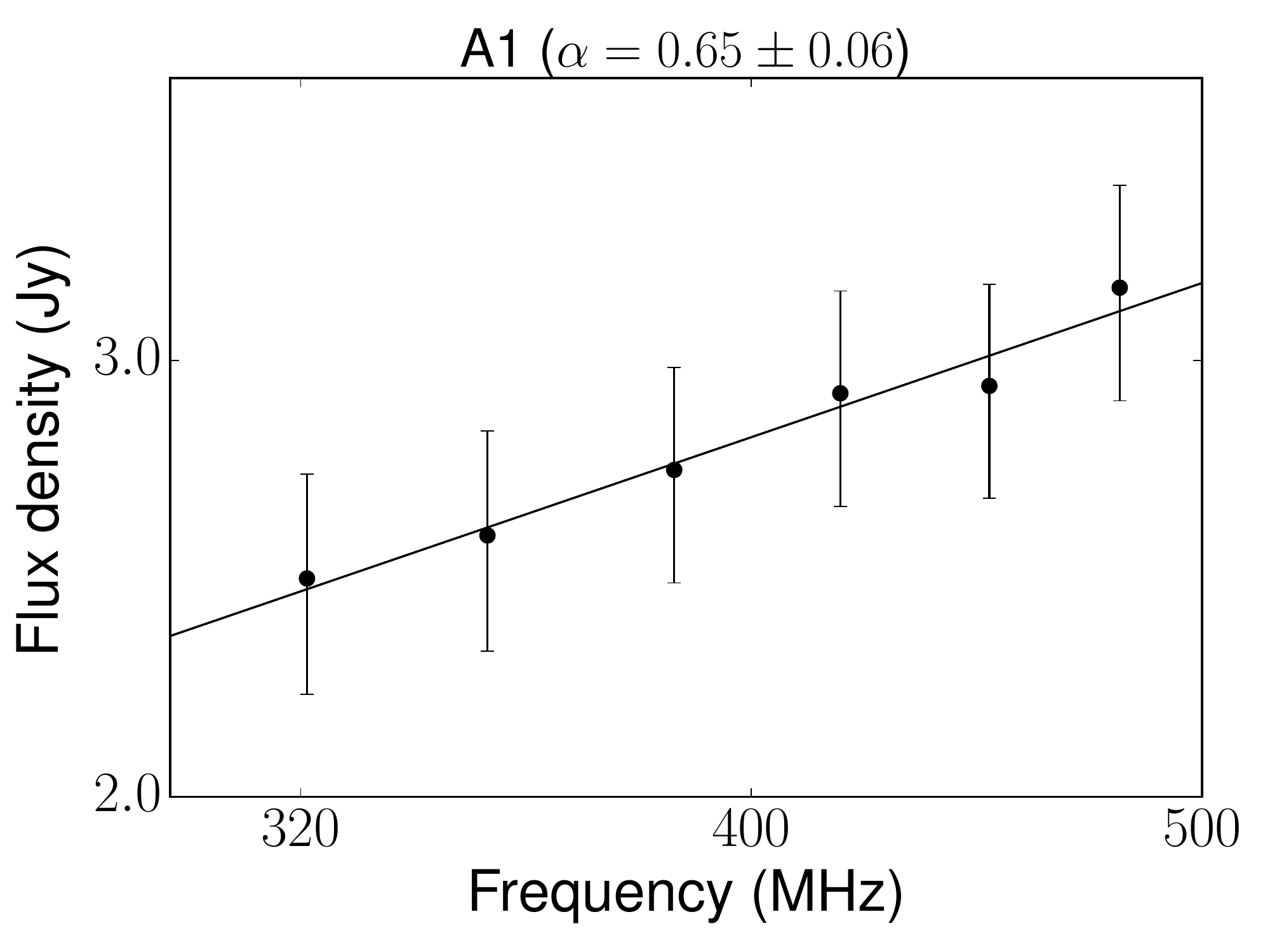} \quad \includegraphics[scale=0.27]{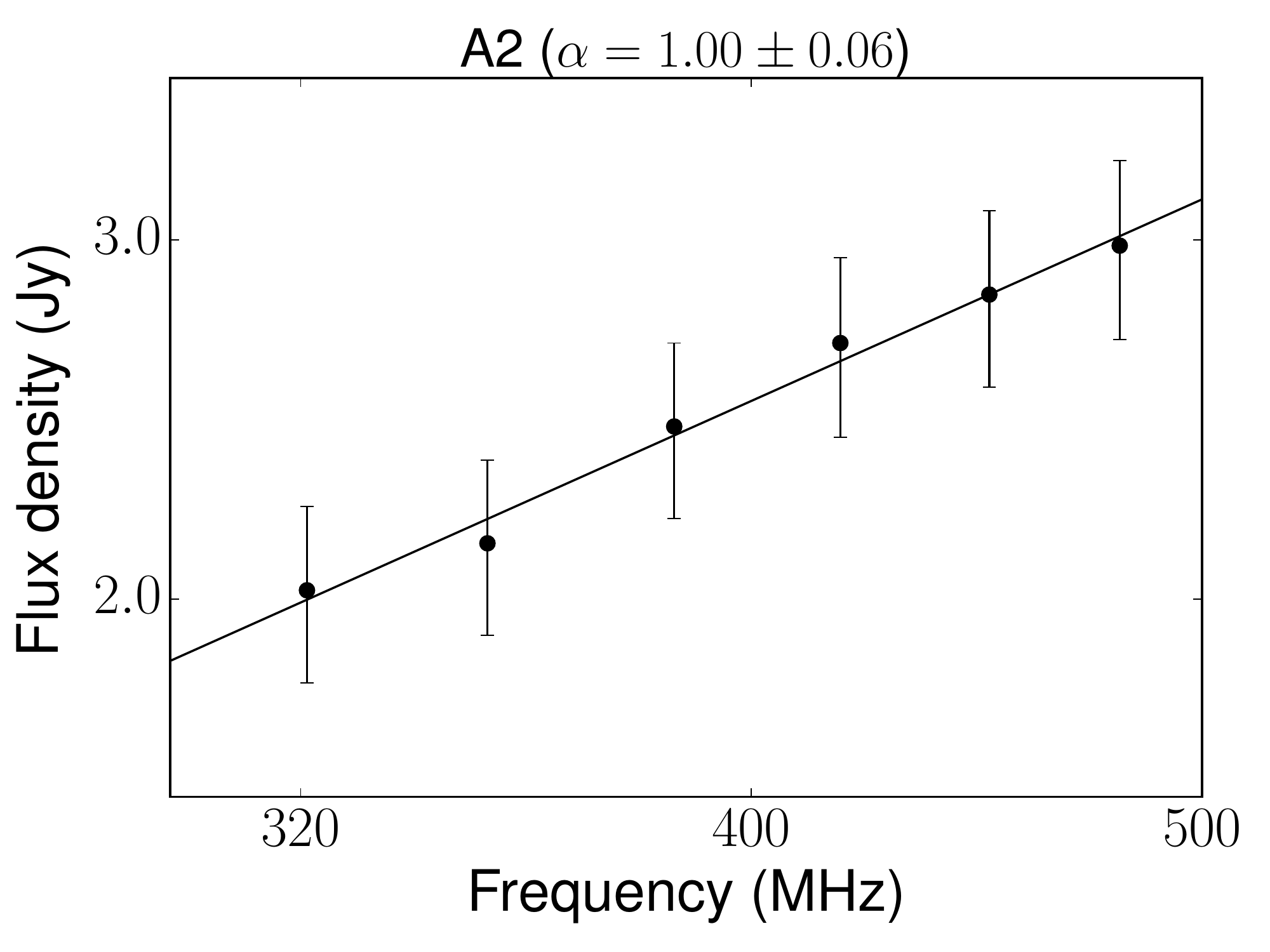} \quad \includegraphics[scale=0.27]{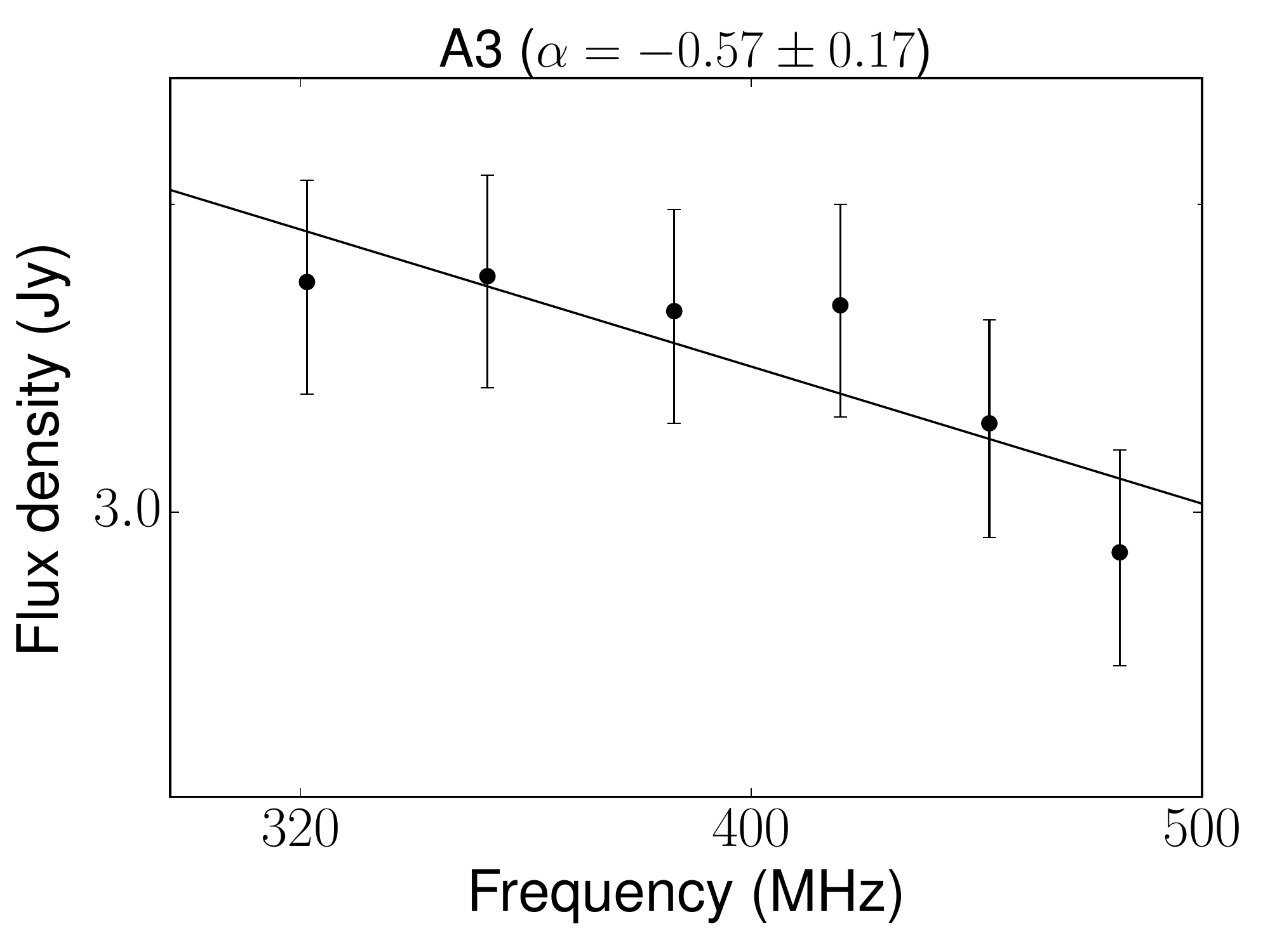} \quad \includegraphics[scale=0.27]{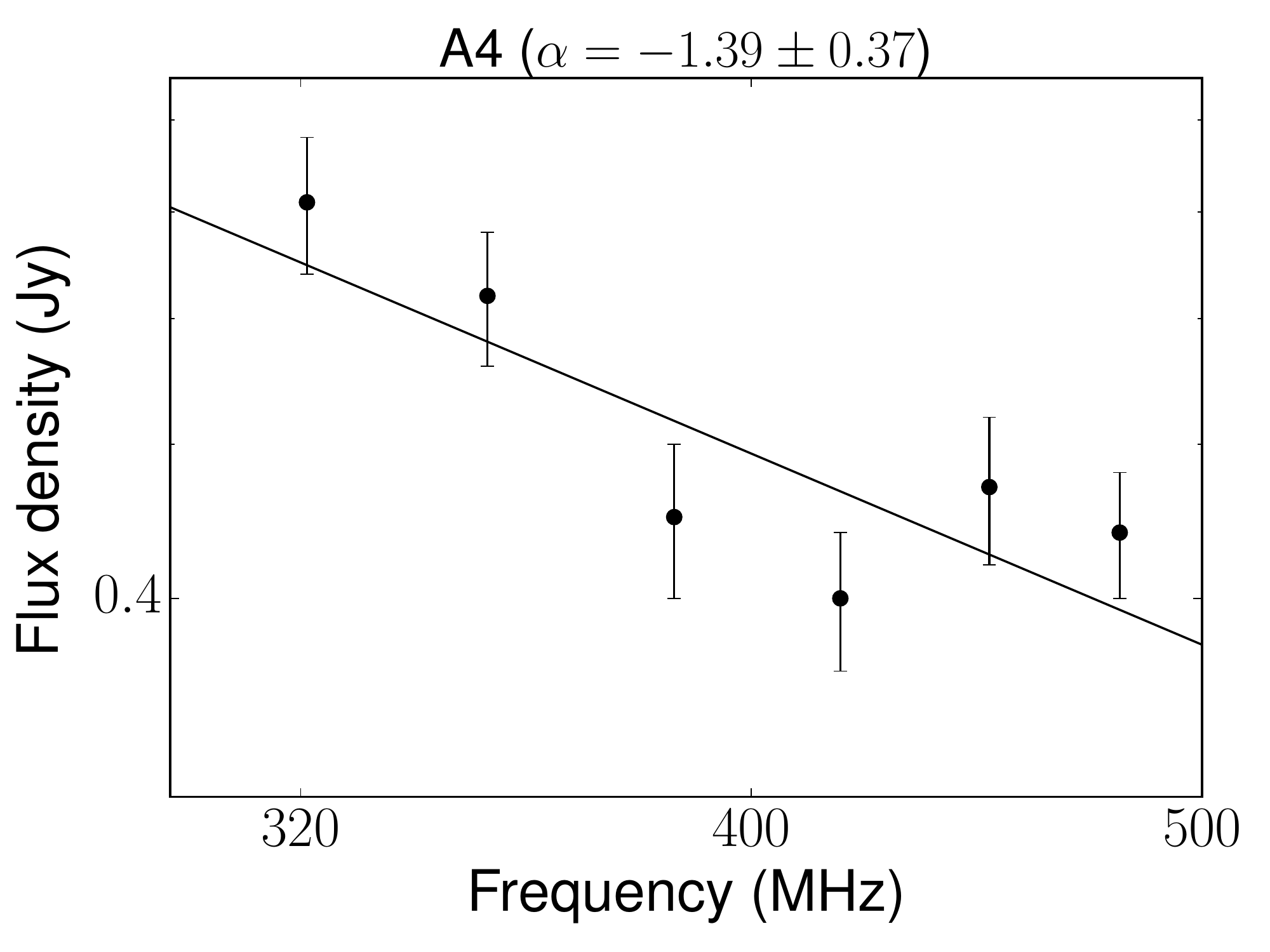} \quad \includegraphics[scale=0.27]{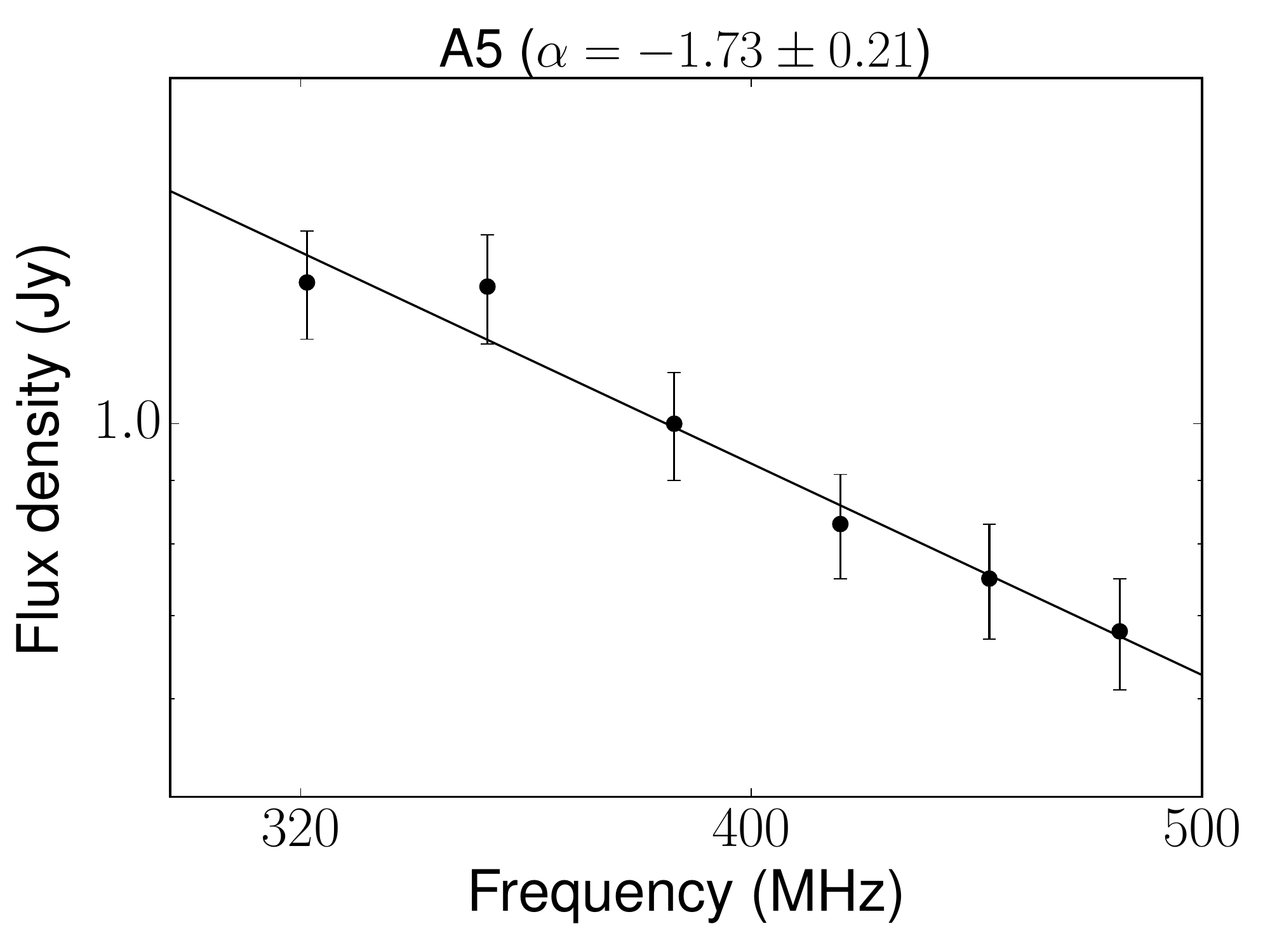} \quad \includegraphics[scale=0.27]{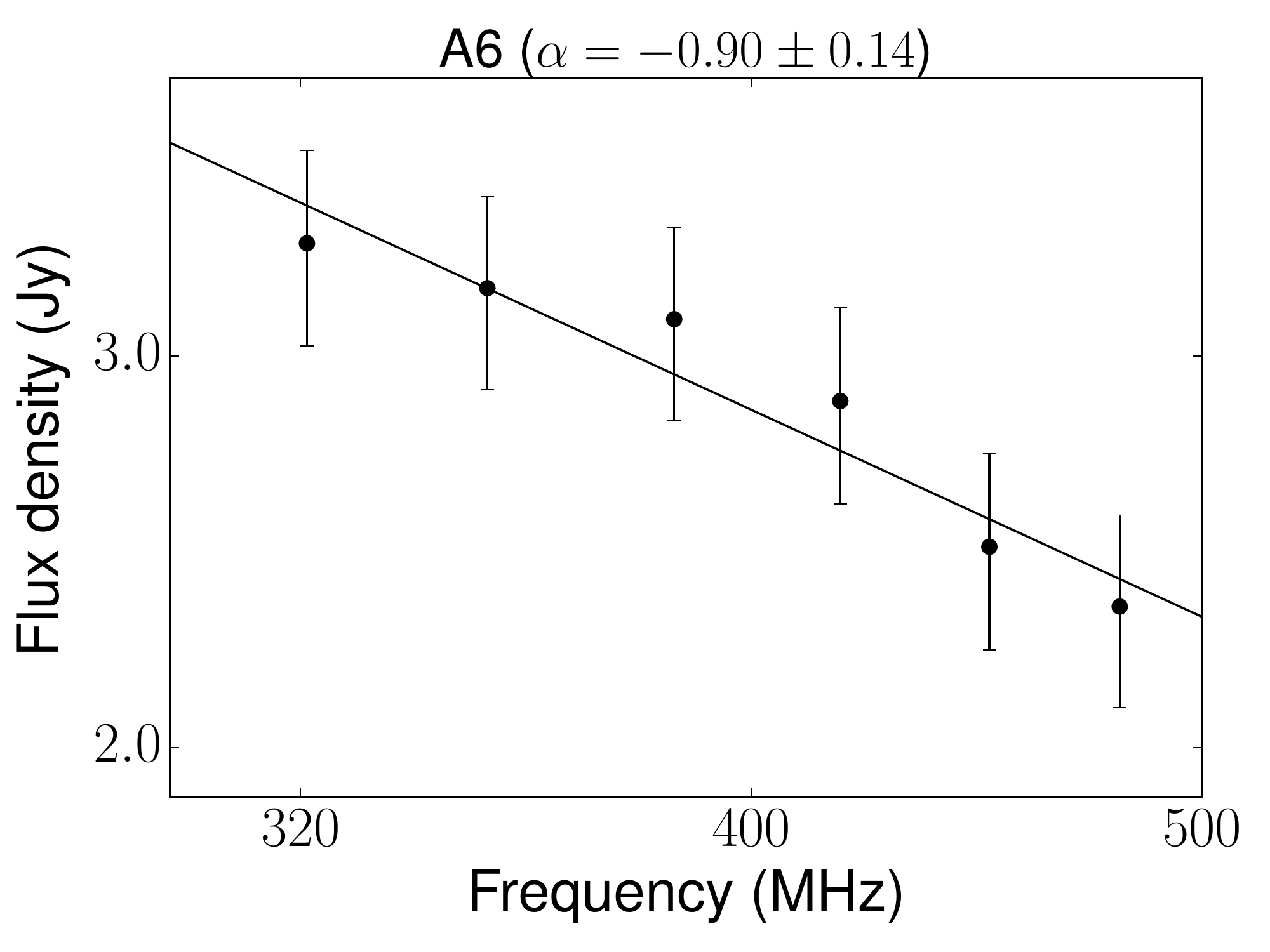} \quad \includegraphics[scale=0.27]{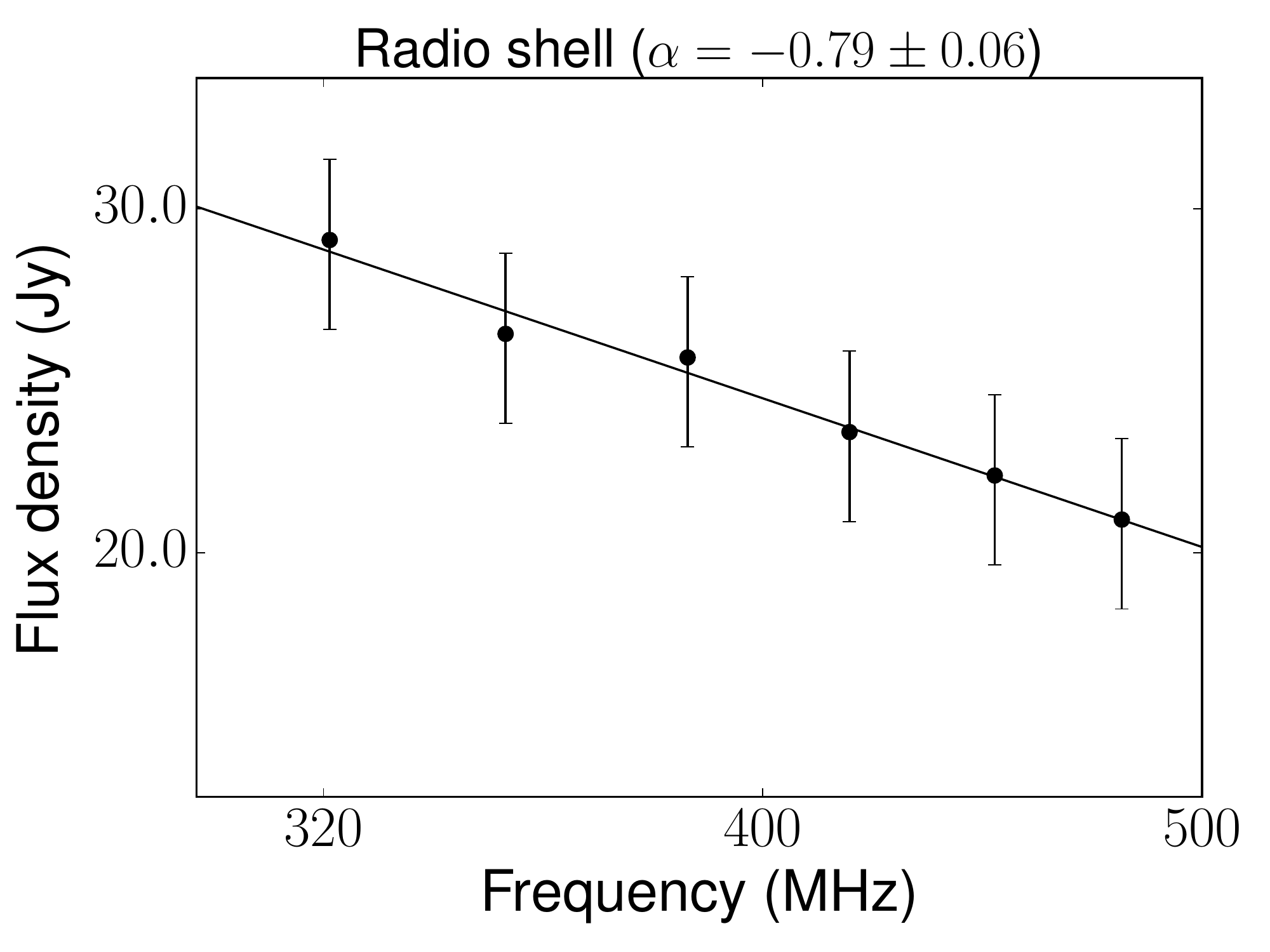}
\caption{Radio spectra of the 6 apertures and the full radio shell using fluxes at 321, 351, 385, 418, 450 and 480~MHz. The corresponding apertures are shown in Fig.~\ref{ugmrt}. }
\label{spcap}
\end{figure*} 

\begin{table*}
\footnotesize
\caption{Flux densities and spectral indices of 6 apertures estimated from least square fit to the fluxes at 6 frequency bands.}
\begin{center}
\begin{tabular}{ c c c c c c c c}\hline 
\setlength{\tabcolsep}{1pt}
Aperture &321~MHz &351~MHz &385~MHz &418~MHz &450~MHz &480~MHz & Spectral index ($\alpha$)\\
 &F(Jy) &F(Jy) &F(Jy) &F(Jy) &F(Jy) &F(Jy) & \\
\hline 
A1 (IRAS 17256--3631)& $2.45\pm0.25$ & $2.55\pm0.26$ & $2.71\pm0.27$ & $2.91\pm0.29$ & $2.93\pm0.29$ & $3.21\pm0.32$ &$0.65\pm0.06$\\
A2 (IRAS 17258--3637)& $2.02\pm0.20$ & $2.13\pm0.21$ & $2.43\pm0.24$ & $2.67\pm0.27$ & $2.82\pm0.28$ & $2.98\pm0.30$ &$1.00\pm0.06$\\
A3& $3.72\pm0.37$ & $3.74\pm0.37$ & $3.62\pm0.36$ & $3.64\pm0.36$ & $3.26\pm0.33$ & $2.89\pm0.29$ &$-0.57\pm0.17$\\
A4& $0.71\pm0.07$ & $0.62\pm0.06$ & $0.45\pm0.05$ & $0.40\pm0.04$ & $0.47\pm0.05$ & $0.44\pm0.04$ &$-1.39\pm0.37$\\
A5&$1.30\pm0.13$ & $1.29\pm0.13$ & $1.00\pm0.10$ & $0.83\pm0.08$ & $0.75\pm0.08$ & $0.68\pm0.07$ &$-1.73\pm0.21$\\
A6&$3.37\pm0.34$ & $3.22\pm0.32$ & $3.12\pm0.31$ & $2.86\pm0.29$ & $2.46\pm0.25$ & $2.31\pm0.23$ &$-0.90\pm0.14$\\
Full radio shell$^\star$&$28.92\pm2.89$ & $25.89\pm2.59$ & $25.18\pm2.52$ & $23.06\pm2.31$ & $21.91\pm2.19$ & $20.80\pm2.08$ &$-0.79\pm0.06$\\
\hline\\
\multicolumn{6}{l}{%
\begin{minipage}{9.5cm}
\scriptsize{$^\star$ : Spectral index of the diffuse radio shell excluding emission from the two \hii~regions
}  
\end{minipage}%
}
\end{tabular}
\label{snrtb}
\end{center}
\end{table*}

\begin{figure*}
\centering
\hspace*{-1.0cm}
\includegraphics[scale=0.55]{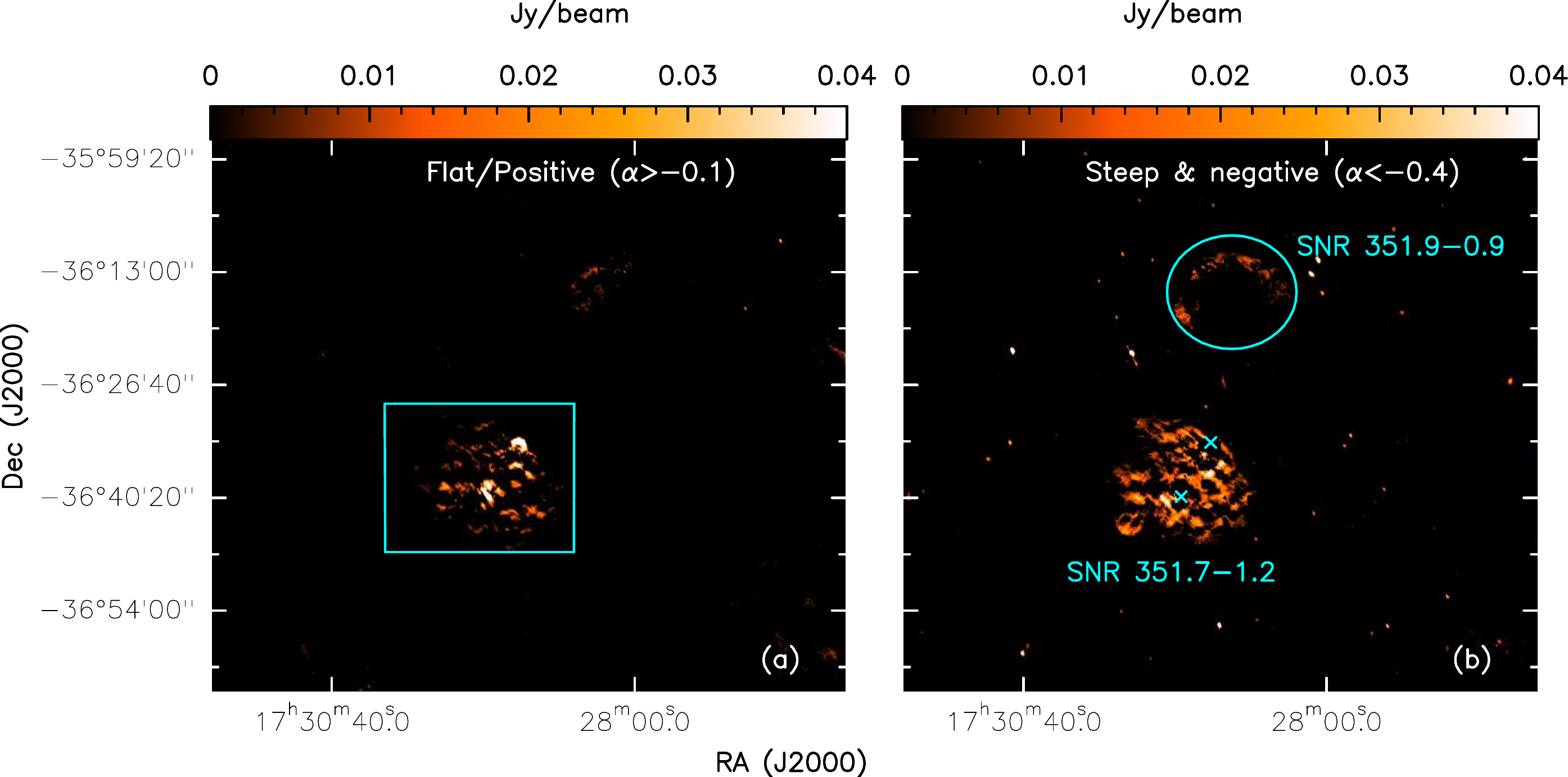}\quad \hspace*{-1.0cm}\includegraphics[scale=0.55]{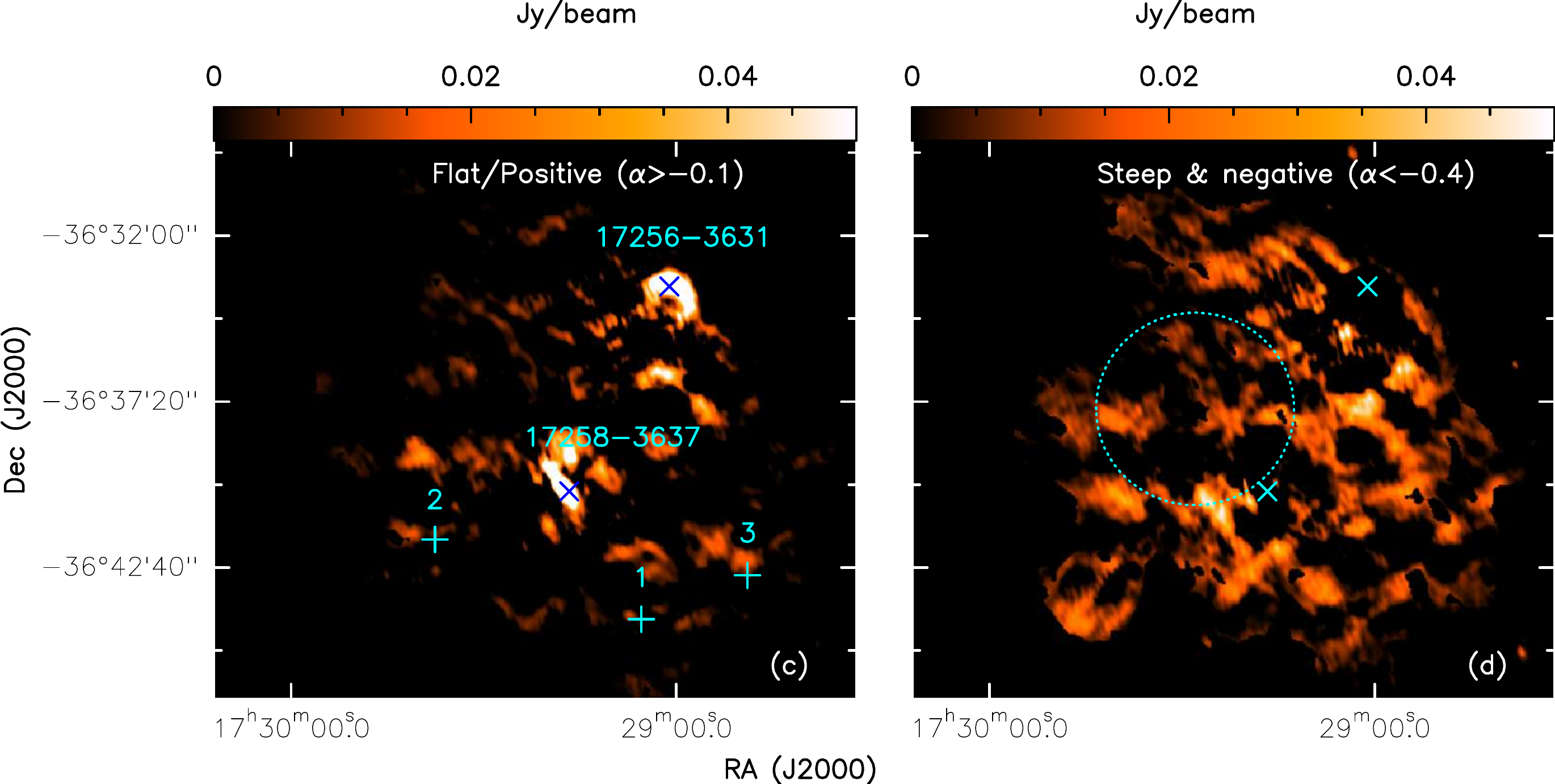}
\caption{Spectral tomography images between 321 and 480~MHz. (a) Intensity scale is reversed to show bright areas that have spectral indices flatter than $-0.1$. (b) Tomography map showing regions whose spectral indices are steeper than $-0.4$, which is predominantly non-thermal. Ellipse shows the location of the SNR G351.9--0.9 (Discussed in Sect.3.8). Panels (c) and (d) show the zoomed tomography images of G351.7--1.2 indicated by a box in panel (a). The locations of the two \hii~regions IRAS 17256--3631 and IRAS 17258--3637 are marked as $\times$ and the approximate location of H$\alpha$ shell is shown as a dotted circle. The peak positions of 3 \hii~regions from \citet{2014ApJS..212....1A} are shown as $+$ in panel (c) and are labelled as 1, 2 and 3.}
\label{tomoful}
\end{figure*}

\subsection{Spectral index}

\par The spectral index analysis is a useful tool to probe the processes responsible for radio emission. The spectral index $\alpha$ is defined as $S_\nu$=$\nu^\alpha$, where $S_\nu$ is the flux density at frequency $\nu$. The radiation processes in \hii~regions are dominated by thermal bremsstrahlung  and hence, they are often referred to as thermal sources. For a homoegenous, spherically symmetric HII region, the flux density due to thermal bremsstrahlung radiation is expected to increase with frequency $\nu$ ($\alpha=2$) until a turnover frequency $\nu_c$ after which, the flux is nearly constant  ($\alpha=-0.1$).  $\nu_c$ represents the turnover frequency where the optical depth is unity. At frequencies below $\nu_c$, the emission is optically thick while it is optically thin above $\nu_c$, that corresponds to flat spectral indices. \citet{1993RMxAA..25...23R} have shown that for thermal free-free emission, spectral index will always be $\alpha\geq-0.1$ irrespective of the source characteristics such as temperature and electron density. Steep negative spectral indices ($\alpha<-0.4$), on the other hand, are often indicative of non-thermal processes \citep{{1988A&A...198..109I},{1999ApJ...527..154K}}. 

\par In order to examine the nature of the radio emission, we estimated the spectral indices towards six apertures shown in Fig.~\ref{ugmrt}. Two of these elliptical apertures encompass the \hii~regions and others are arbitrarily chosen in regions of the shell, where there is bright diffuse emission. We determined the flux densities in the apertures at all the six frequencies and fitted their radio SEDs with the power-law,  $S_\nu$=$\nu^\alpha$, by the least-square fitting method (see Figure~\ref{spcap}). The errors in the flux densities are estimated using the expression $\sqrt{(2\sigma\sqrt{\theta_{src}/\theta_{bm}})^2+(2\sigma')^2}$ where $\sigma$ is the rms noise level of the map,  $\sigma'$ is the the error in flux scale calibration, $\theta_{bm}$ represents the size of the beam, and $\theta_{src}$ is the source size, taken as the geometric mean of the major and minor axes of the elliptical sperture \citep{2013ApJ...766..114S}. The uncertainty in the flux calibration of GMRT is taken to be 5\%  \citep{2007MNRAS.374.1085L}. The spectral indices obtained towards the apertures are listed in Table~\ref{snrtb}. Towards apertures A1 and A2, that correspond to IRAS 17256--3631 and IRAS 17258--3637, the spectral indices are positive signifying thermal free-free emission. However, towards apertures A3, A4, A5 and A6, the spectral indices are steep and negative ($<-0.5$) confirming the presence of non-thermal emission. We have also evaluated the spectral index of the full radio shell (centered on $\rm{\alpha_{J2000}}$: 17$^h$29$^m$18.5$^s$, $\rm{\delta_{J2000}}$: -36$^\circ$37$'$22.0$''$) excluding the two \hii~regions, and the value is $-0.79\pm0.06$, consistent with non-thermal emission. A number of studies have reported similar spectral indices in shell-like SNRs \citep[$\alpha\sim-0.6~\textrm{to}-0.8$;][]{{1998MNRAS.296..813G},{2001A&A...366.1047C},{2014BASI...42...47G}}. The morphology as well as the spectral indices suggest that this large scale non-thermal emission is plausibly related to a previously unknown SNR candidate G351.7--1.2 (hereafter SNR~G351.7). 

We have also compared the GMRT flux densities with the values at 843~MHz obtained from SUMSS survey \citep{1999AJ....117.1578B} to investigate whether the estimated spectral indices conform with the higher frequency SUMSS flux densities. We find that the SUMSS flux densities are lower than GMRT flux values in all apertures. A direct comparison of spectral indices and flux densities is difficult as (i) the spatial scales of emission (i.e. visibility ranges of GMRT and SUMSS) are not identical, and (ii) the sensitivities are different (1 mJy/beam for GMRT and 5 mJy/beam for SUMSS, where the beam size of SUMSS image is $58\arcsec$).

\par The spectral indices of emission in apertures A4 and A5 are $-1.2$ and $-1.5$, respectively. These are steeper than the value of the mean spectral index ($\alpha\sim-0.8$) of the radio shell. Steep spectral indices ($\alpha\lesssim-0.5)$ are expected in regions where the effect of synchrotron losses are significant \citep{2009ApJ...703..662R}. Steep spectral indices are also observed in regions where shocks are decelerated due to collision with dense clouds as seen in the case of Puppis A \citep{1991AJ....101.1466D}. 
A close scrutiny of the radio SEDs shows that most of them are in accordance with a power-law. However, some of them (A3, A4 and A6) also show hints of curvature in spectra. The curvature in radio spectra is reminiscent to those observed towards many Galactic and extragalactic SNRs \citep[e.g.,][]{{2008SerAJ.177...61C},{2012ApJ...756...61O}}, but over a larger frequency range (few GHz) unlike our case ($\sim500$~MHz). In the present work, we abstain from analysing and interpreting the radio curvature. The variation of spectral indices as well as radio spectra along different apertures point towards the complexity of this region. 

Based on the radio continuum data and WISE mid-infrared emission, \citet{2014ApJS..212....1A} have identified few \hii~regions in this region including IRAS 17256--3631 and IRAS 17258--3637. We have also calculated the spectral indices of 3 of these \hii~regions (labelled as 1, 2 and 3 in Fig.~\ref{tomoful}) whose positions overlap with the diffuse radio shell. The spectral indices of these regions are found to be $-0.3\pm0.3,0.2\pm0.3$ and $0.3\pm0.2$, respectively. These values are consistent with thermal free-free emission within the uncertainties. At low radio frequencies ($\nu<1$~GHz), the thermal emission is expected to be optically thick ($\alpha\sim2$). The flatness of the obtained spectral indices can be interpreted as a combined effect of thermal emission from the \hii~regions and non-thermal emission from the radio shell.

\subsubsection{Spectral tomography}
We also use an additional technique called the spectral tomography to probe the emission from SNR~G351.7. This technique is beneficial in regions where the emission consists of overlapping thermal and non-thermal structures. In our case, these correspond to the \hii~regions as well as radio emission from the SNR. Initially used to analyse the spectra of radio galaxies \citep{1997ApJ...488..146K}, this method has also been applied to various SNRs \citep{{2000ApJ...529..453K},{2002ApJ...580..914D}}. The tomography method involves the construction of a set of tomography images in which intensity images are created according to the following expression:

\begin{equation}
I_t(\alpha_t) \equiv I_{321} - \left(\frac{321}{480}\right)^{\alpha_t}I_{480}
\end{equation} 

\noindent Here, $I_t(\alpha_t)$ is the tomographic image corresponding to the threshold spectral index $\alpha_t$. $I_{321}$ and $I_{480}$ are the images at 321 and 480~MHz respectively. If a component has spectral index $\alpha_t$, it gets subtracted from the image $I_t$. If a component has spectral index greater than the threshold spectral index, it will be oversubtracted from the images whereas regions having spectral indices lower than $\alpha_t$ will be undersubtracted and will thus have positive brightness. We have applied this method to our region of interest. In constructing the tomographic images, we have only considered those pixels whose intensity is greater than 3$\sigma$, where $\sigma$ is the rms noise of the image.  

\begin{figure}
\centering
\hspace*{-0.8cm}
\includegraphics[scale=0.31]{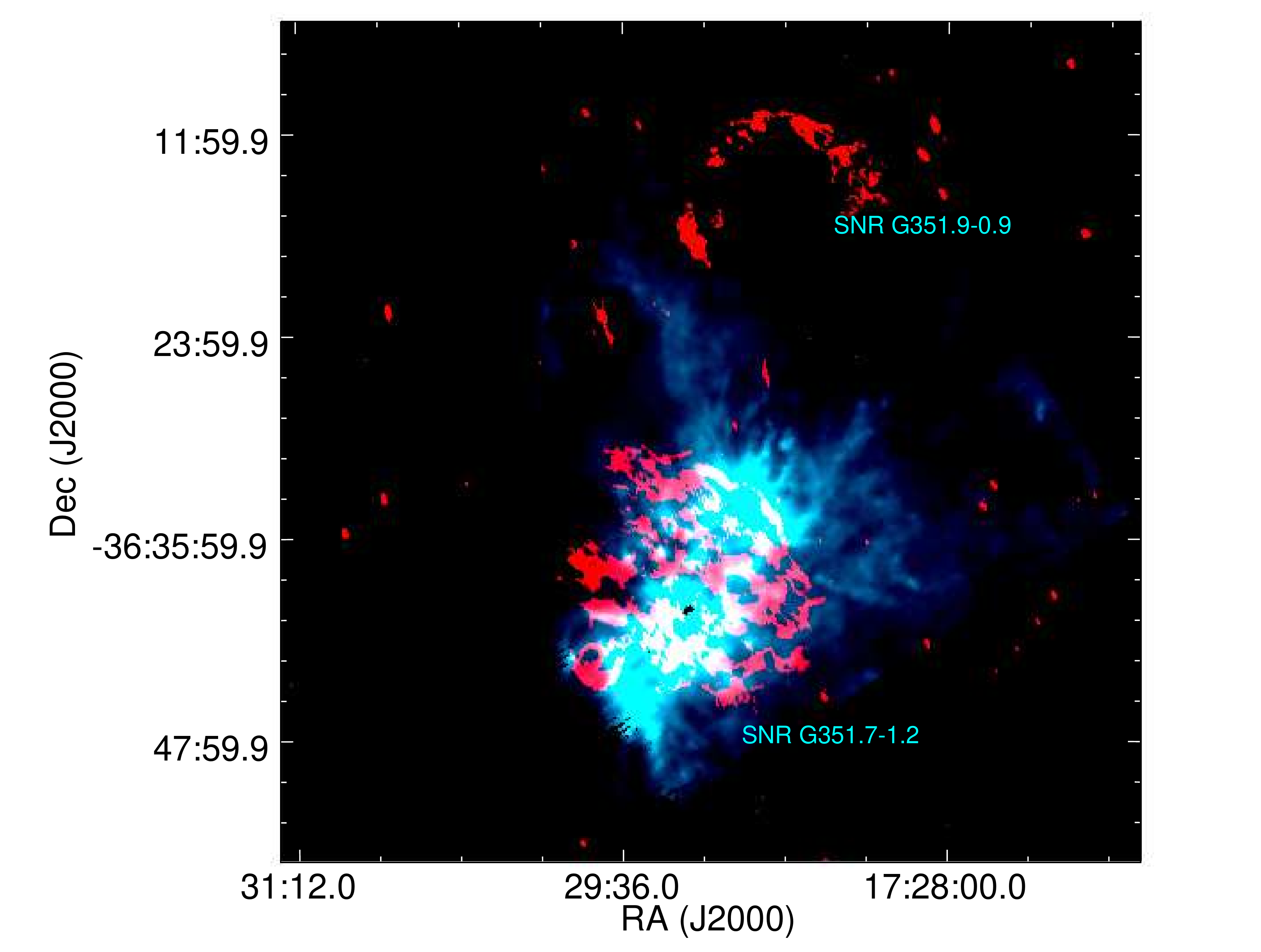}
\caption{Colour composite image of tomography map with negative spectral indices steeper than $-0.4$ (red) and cold dust emission at 250~$\mu$m from Herschel SPIRE (cyan). The SNR candidate G351.7-1.2 and SNR  G351.9-0.9 are also labelled.}
\label{tom_250}
\end{figure}

\par We have considered regions with $\alpha_t>-0.1$ to be those associated with thermal emission. For inferring about non-thermal emission, we have constructed a tomographic image with $\alpha_t<-0.4$. Locations with spectral indices $-0.4<\alpha<-0.1$ are assumed to have contributions from thermal as well as non-thermal emission. The tomographic map showing thermal emission is displayed in panel (a) of Fig.~\ref{tomoful}. The color scale is inverted so that the regions whose spectral indices are more positive than $-0.1$ are seen as bright emission.  Prominent features in the image include the two \hii~regions, visible in panel (c) of Fig.~\ref{tomoful}. Panels (b) and (d) of Figs.~\ref{tomoful} refer to non-thermal tomographic maps constructed using a threshold spectral index $\alpha_t=-0.4$, thus pointing towards regions where synchrotron emission dominates. Unlike apertures, the analysis is carried out at the level of each pixel, so that the small scale variations are prominent. We clearly discern a spherical distribution in the non-thermal emission map. However, the emission is filamentary in nature that is reminiscent of plerion SNRs. Also limb-brightening, a characteristic of shell SNRs, is not manifested in the map. A plausible reason for the filled filamentary emission could be the presence of molecular clouds in this region. Fig.~\ref{tom_250} shows the colour composite image of G351.7-1.2, where the non-thermal spectral tomography image ($\alpha\leq-0.4$, shown in red) is combined with the image of cold dust emission at 250 ~$\mu$m (cyan). The molecular cloud is seen as high density filamentary structures and most of the steep non-thermal emission is located towards regions of relatively lower density. Researchers have examined local variations of SNR spectral indices (especially flattening) in molecular clouds and several theories have been proposed to understand the emission mechanisms. These include second-order fermi acceleration in the turbulent medium near the shock and momentum diffusion \citep[e.g.,][]{{1989A&A...219..192S},{1999A&A...345..256O}}. 

In the following sections, we provide details of supporting evidence regarding the SNR. We also notice an arc-like feature to the north of the SNR G351.7, marked in Fig.~\ref{tomoful}(b). This prominent shell-like arc has been previously identified as SNR~G351.9--0.9 catalogued by \citet{2014BASI...42...47G} and we discuss more about this SNR later, in Sect. 3.7.

\begin{figure}
\centering
\hspace*{-0.5cm}
\includegraphics[scale=0.32]{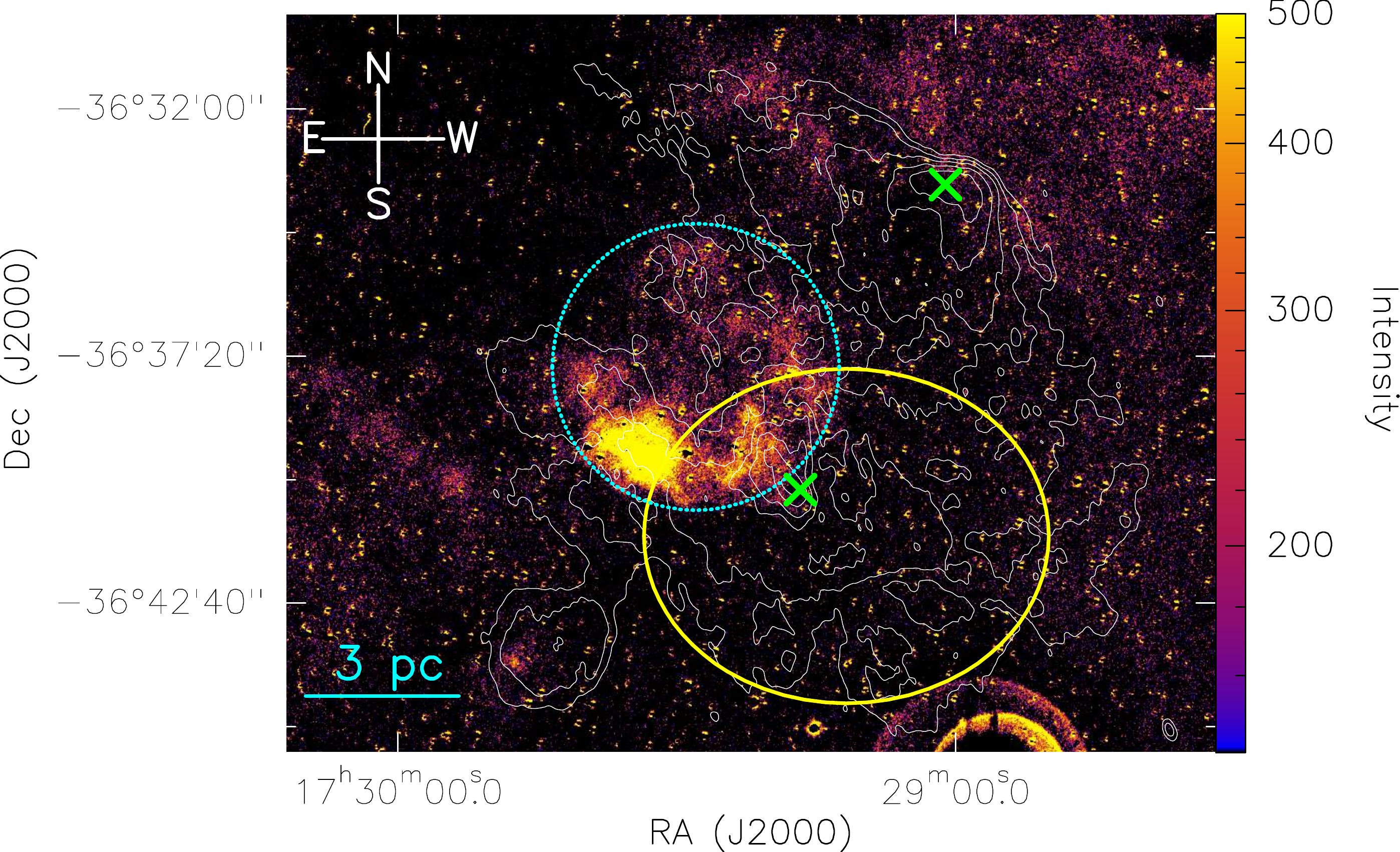}
\caption{H$\alpha$ image of SNRG351.7 overlaid with 321~MHz radio contours. Contour levels are 15, 36, 75, 150, 250 and 350~mJy/beam. The peak positions of two \hii~regions IRAS 17256--3631 and IRAS 17258--3637 are marked as $\times$. Dotted circle denotes the approximate location of H$\alpha$ shell. Ellipse denotes the location of $Fermi$-LAT $\gamma$-ray source (95$\%$) confidence level.}
\label{halpha}
\end{figure}  

\subsection{Optical emission towards G351.7}

We gathered that the observed radio shell is a SNR candidate from the previous subsections. Observational studies show that $\sim30$ percent of the Galactic SNRs are associated with optical emission \citep{2014BASI...42...47G}. As the SNR blast wave propagates into the cool and dense ambient ISM, shock excited radiation can develop at the interface. This is observed as H$\alpha$, other Balmer lines as well as forbidden transitions such as [SII], [NII], [OIII] etc. In addition to the shocked interface, strong optical emission is also observed in cooling and recombination zones posterior to the shock itself \citep[e.g.,][]{{2000AJ....120.1933D},{2010ApJ...710..964D}}. These are the regions where inelastic collisional processes cool the gas to lower temperatures. 
In the quest for more evidence, we have searched for a possible counterpart to the radio emission in the optical wavebands towards the SNR candidate G351.7. We utilized the SuperCOSMOS H-alpha survey image for this purpose and subtracted the continuum emission using a scaled R-band image from the same survey. The H$\alpha$ image displays considerable optical emission in the form of a shell-like feature. The optical H$\alpha$ emission towards SNR~G351.7 is shown in Fig.~\ref{halpha}, overlaid with contours of radio emission at 321~MHz. The H$\alpha$ shell overlaps with the inner edge of the radio shell.  A striking feature is the  opening towards the north-east, in the same direction as the gap in the radio emission. The diameter of the H$\alpha$ shell is  $\sim5.6\arcmin$, that corresponds to 3.3 pc for a distance of 2~kpc. The H$\alpha$ brightness is not uniform across the rim of the shell. It is brighter towards the south-east and regions of low brightness across the shell appear filamentary. We also notice that H$\alpha$ emission is not detected in the regions where the radio brightness of the two \hii~regions peak. The SNR and \hii~regions could possibly be located within the molecular cloud. If that is the case, the weak optical emission could be attributed to effects of extinction. This is substantiated by high extinction found in this region \citep[$10-30$~mag;][]{{2014MNRAS.440.3078V},{2016MNRAS.456.2425V}}.

\par H$\alpha$ emission encompassed by large scale radio emission has been observed in other Galactic SNRs \citep[e.g.,][]{{1997AJ....113.1379G},{2009MNRAS.394.1791S}}. One archetypal example is that of SNR W28 \citep{2000AJ....120.1933D}. The radio emission associated with this SNR is diffuse and possesses a shell morphology and the H$\alpha$ emission is found towards the interior side of the radio shell-like emission.  This SNR is similar to G351.7-1.2 in many aspects as it is associated with a molecular cloud and also harbours many HII regions.

\subsection{Infrared emission towards G351.7} 

\begin{figure*}
\centering
\hspace*{-0.5cm}
\includegraphics[scale=0.6]{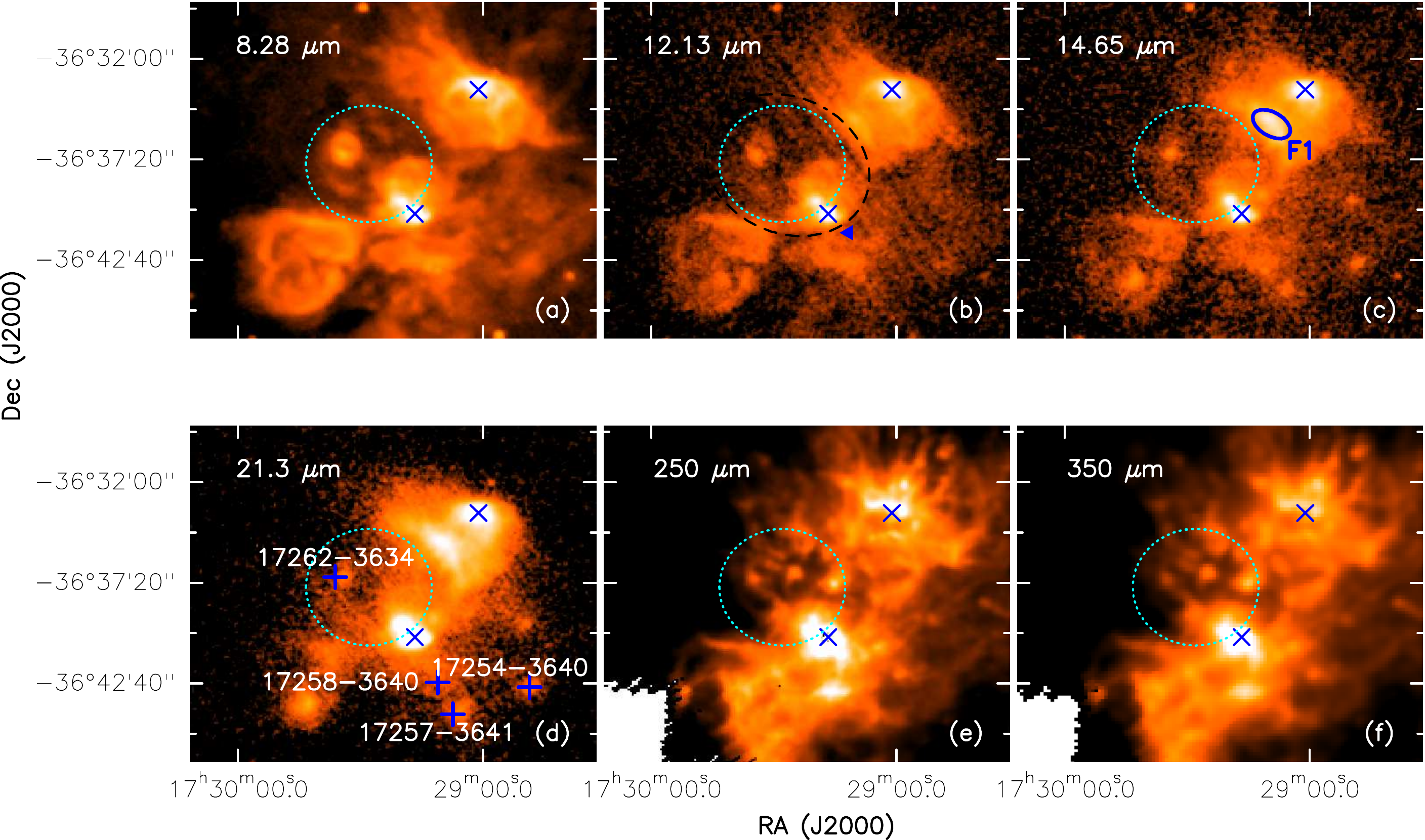}
\caption{Multiwavelength view of SNR~G351.7 and its surroundings. Panels (a), (b), (c) and (d) corresponds to mid-infrared emission from MSX at 8.28, 12.13, 14.65, and 21.3~$\mu$m. Panels (e) and (f) shows cold dust emission mapped using $Herschel$ at 250 and 350~$\mu$m. Dotted circle denotes the approximate location of H$\alpha$ shell. Triangle in panel (b) shows the location of the $\gamma$-ray source. Ellipse in panel (c) shows the bright arc-like feature F1 seen in the periphery of H$\alpha$ shell. Other IRAS sources in the field are marked as crosses in panel (d).}
\label{msx}
\end{figure*}  

Using mid-infrared and submillimeter data, we have investigated the dust environment of SNR~G351.7. Fig.~\ref{msx} shows the emission at six wavebands from 8.28 to 350~$\mu$m. The mid-infrared maps (from \textit{Spitzer}-IRAC and MSX) reveal two bright sources that correspond to the \hii~regions IRAS 17256--3631 and IRAS 17258--3637. In addition, emission associated with four other IRAS sources is also discerned. These are marked in Fig.~\ref{msx}(d). We interpret the mid-infrared emission from these sources as tracing the ongoing star forming activity. The 12.13, 14.65 and 21.3~$\mu$m MSX maps (see Fig.~\ref{msx}) show the presence of bright, arc-like features, highlighted by a dashed curve in panel (b) of Fig.~\ref{msx}. These lie on the outer periphery of the H$\alpha$ shell (represented as a circle in Fig.~\ref{msx}) and we construe these as shock heated dust plausibly swept up by the supernova blast wave \citep{2012MNRAS.420.3557G}. The fact that these features are seen prominently in mid-infrared ($12-21$~$\mu$m) could be related to the size of dust grains that survive the blast-wave and are shock-heated \citep[e.g.,][]{{2006ApJ...652L..33W},{2010ApJ...725..585A}}. We again distinguish the lack of infrared emission towards the north-east.

\par Cold dust emission associated with molecular clouds is clearly perceived using the longer wavelength far-infrared bands. The regions of bright emission correspond to the two \hii~regions, with diffuse emission observed prominently to the west and south-western sides of the H$\alpha$ emission. There is no noticeable emission towards the eastern and north-eastern sides of the SNR candidate. The shock-heated arc-like features perceived in mid-infrared are located between the outer edge of the H$\alpha$ emission and the inner regions of cold dust emission. This suggests strong interaction of the SNR with the surrounding high density cloud. 

\begin{figure}
\centering
\hspace*{-0.5cm}
\includegraphics[scale=0.45]{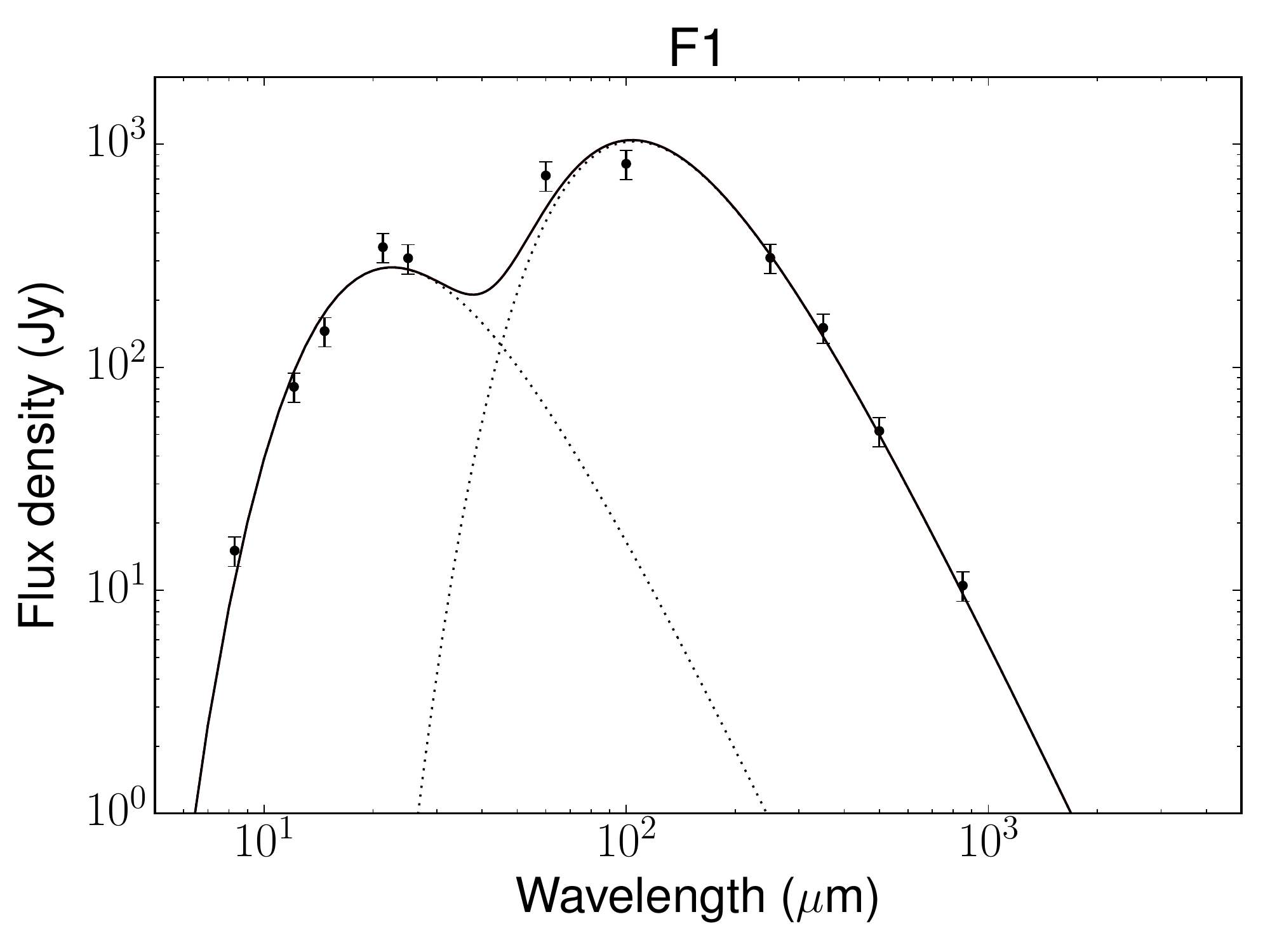}
\caption{The infrared-submillimeter ($8.3-850~\mu$m) SED including MSX, IRAS, $Herschel$ and Apex+Planck towards F1. The two temperature blackbody fit to the thermal component is shown as solid black line.}
\label{sedf1}
\end{figure}

\begin{figure*}
\centering
\hspace*{0cm}
\includegraphics[scale=0.4]{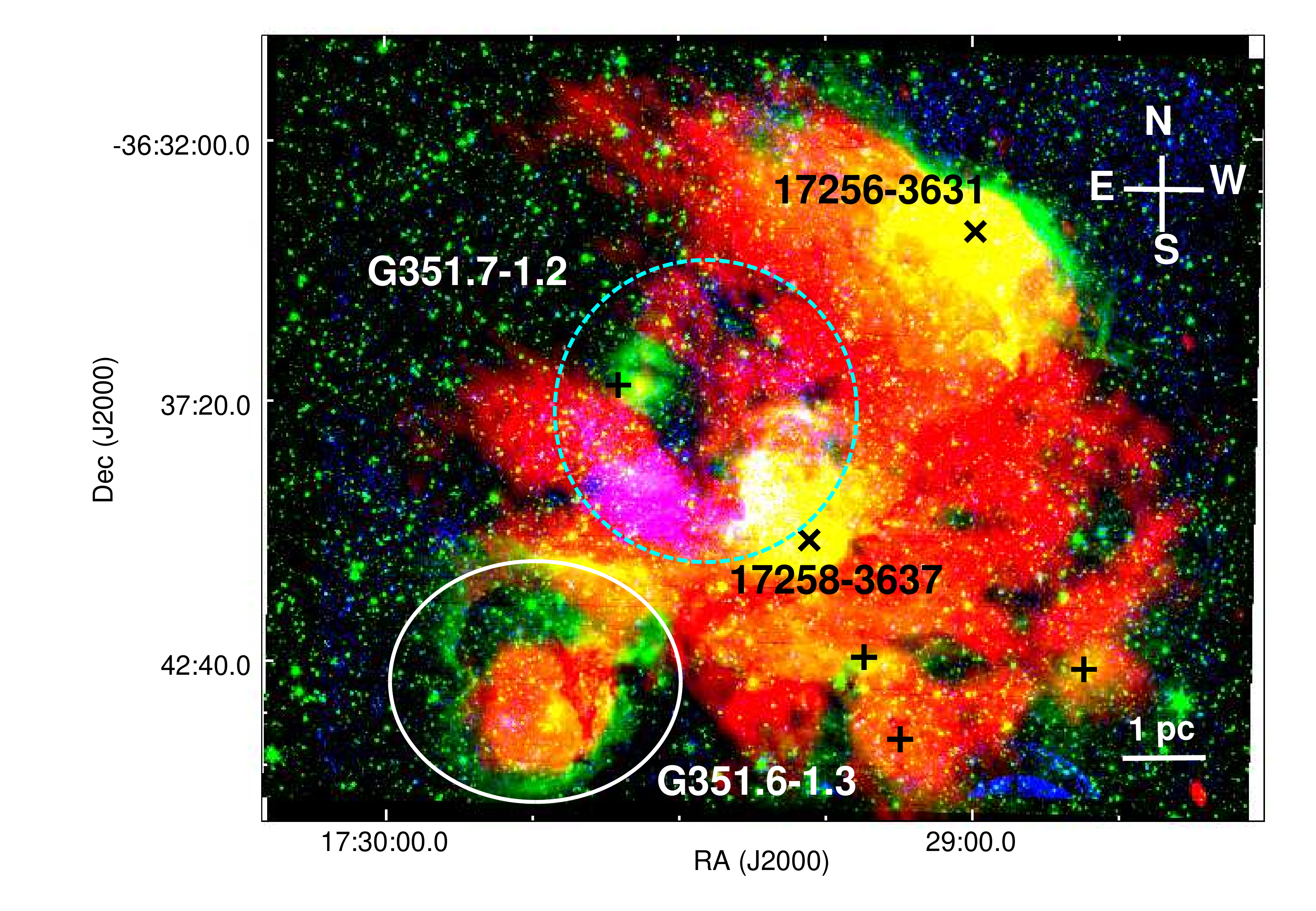}
\caption{Three color composite image of the SNR~G351.7--1.2. The 321~MHz radio map is shown in red, 4.5~$\mu$m infrared map in green and optical (H$\alpha$) image in blue. The two \hii~regions IRAS 17256--3631 and IRAS 17258--3637 marked are labelled. The SNR shell is evident in the image. The H$\alpha$ shell (dotted circle) and the bubble G351.6--1.3 (solid ellipse, discussed in Sect. 3.7) are also marked. The locations of the other IRAS sources in the field of view are indicated with +.}
\label{colcom}
\end{figure*}  

\subsubsection{Spectral Energy Distribution towards F1}

\par We next estimate the temperature of the warm dust in the arc-like structures. We select the arc-like feature F1 (shown in panel (c) of Fig.~\ref{msx}) and present its infrared-submillimeter SED. The flux densities from MSX, IRAS, $Herschel$ and ATLASGAL+Planck at 11 wavebands (8.28, 12.13, 14.65, 21.3, 25, 60, 100, 250, 350, 500 and 850~$\mu$m) are determined by integrating the intensities within an elliptical aperture of size $124\arcsec\times74\arcsec$ oriented along the arc. The SED  is displayed in Fig.~\ref{sedf1} and exhibits characteristics of two dust components. Consequently, the flux densities in the SED are fitted with a function of the form shown in Eqn. (2). This equation is the outcome of radiative transfer through a homogeneous absorbing medium comprised of two dust components, reflected in the two-component modified blackbody functions corresponding to temperatures $T_\textrm{1}$ and $T_\textrm{2}$.

\begin{equation}
F_{\nu}=\Omega\,[a\,B_{\nu}(T_1)+(1-a)\,B_{\nu}(T_2)]\,(1-e^{-\tau_\nu}) 
\label{bbody}
\end{equation}

\noindent where

\begin{equation}
\tau_\nu=\mu\  m_\textrm{H} \, \kappa_{\nu} \, N(\textrm{H}_2)
\label{tau}
\end{equation}

\noindent Here, $\Omega$ is the solid angle subtended by the clump, $a$ is the ratio of the warm dust emission to the total emission, $B_{\nu}(T_\textrm{1})$ is the blackbody function at dust temperature $T_\textrm{1}$ and $B_{\nu}(T_\textrm{2})$ is the blackbody function at dust temperature $T_\textrm{2}$, $\mu$ is the mean weight of molecular gas taken to be 2.86, assuming that the gas is 70$\%$ molecular hydrogen by mass \citep{2010A&A...518L..92W}, $m_\textrm{H}$ is the mass of hydrogen atom, $\kappa_\nu$ represents the dust opacity, and $N(\textrm{H}_2)$ is the molecular hydrogen column density. The dust opacity is estimated using the expression \citep{2010A&A...518L..92W}, 

\begin{equation}
\kappa_{\nu}=0.1\left(\frac{\nu}{1000\rm{GHz}}\right)^{\beta} 
\label{opacity}
\end{equation}

\noindent where $\nu$ is the frequency and $\beta$ is the dust emissivity index. We choose $\beta$ as 1.5 \citep{2012MNRAS.420.3557G}. The best fits were obtained using non linear least squares Marquardt-Levenberg algorithm, considering $T_1$, $T_2$ and $N(\textrm{H}_2)$ as free parameters. We have assumed a flux density uncertainty of 15$\%$ in all bands \citep[][and references therein]{2016MNRAS.456.2425V}. The best fit temperatures in the filament are $T_1=31.6\pm1.9$~K and $T_2=153.1\pm7.9$~K. The presence of warm dust (153~K) is likely to be due to (i) collisional heating by plasma, or (ii) radiative shocks. In the former case, the dust would reside in regions coincident with X-ray emission, while in the latter case, the dust emission would overlap UV/optical knots or filamentary structures \citep{2006ApJ...650..212B}. The lack of optical H$\alpha$ emission in the filamentary region prompts us to consider collisional heating by the plasma as the source of duct heating since radio emission has also been detected in this region.  Assuming collisional heating of dust in the shocked gas, the dust temperature of $\sim150$~K is consistent with a plasma of electron density $\geq300$~cm$^{-3}$ \citep{2006ApJ...650..212B}. However, we note that a lack of optical emission could also be ascribed to extinction due to the cloud in this region. The heating mechanism can be reaffirmed by X-ray observations of this region. 

\subsection{$\gamma$-ray source 1FGLJ1729.1--3641c: Possible association with SNR}
SNRs are widely believed to be a major source of Galactic cosmic rays (GCR). According to \citet{1994A&A...287..959D}, $\gamma$-rays are expected in SNRs that are interacting with molecular clouds and this has been confirmed observationally \citep[e.g.,][]{{2011arXiv1104.1197U},{2015A&ARv..23....3D}}. These $\gamma$-rays are produced as a result of hadronic interactions between cosmic rays and the dense molecular cloud. Towards G351.7, a $\gamma$-ray source 1FGLJ1729.1--3641c has been found that is located towards south-west of the radio shell (see Fig.~\ref{halpha}). The origin of this $\gamma$-ray source, listed in the Fermi LAT catalog, has been speculated to be the interaction between the stellar wind from the massive star in the \hii~region IRAS 17258--3637, and the associated molecular cloud \citet{2015ApJS..217....2M}. A visual inspection towards the region enclosed within the 95\% confidence level of this source location (represented as an ellipse in Fig.~\ref{halpha}), shows diffuse radio emission as well as high extinction filamentary structures. Although the angular resolution ($\sim0.1^\circ$) is insufficient to pinpoint its origin, we propose that the $\gamma$-ray source has a likely genesis at the site where SNR interacts with the cloud. However, we cannot rule out the possibility of a chance association of SNR and $\gamma$-ray source.

\subsection{SNR interaction with the molecular cloud: Is the star-formation triggered by the supernova?}

The multiwavelength portrayal of this region can be visualized from Fig.~\ref{colcom}. In this color-composite image, radio emission is shown as red, \textit{Spitzer} IRAC 4.5~$\mu$m as green and H$\alpha$ emission as blue. In the mid-infrared (i.e. green), the prominent features are the two HII regions and the bubble candidate G351.6-1.3 (discussed in Section 3.8). We also detect weak infrared emission in the shell region. If the large scale radio emission were of thermal origin (HII region/wind blown bubbles),  the infrared emission in the radio shell would have been comparable with that of emission from nearby HII regions. The lack of strong, diffuse mid-infrared emission in the radio shell suggests negligible contribution from free-free emission. Majority of the Galactic SNRs show little or no evidence of dust emission in the IRAC bands. However, \citet{2006AJ....131.1479R} identify a sample of SNRs that are associated with mid-infrared emission. These SNRs are found to interact with dense gas/molecular clouds. This is in broad agreement with our hypothesis that the proposed SNR candidate is interacting with the ambient molecular cloud.

\par The morphology of dust and consequently molecular gas in this region can be deciphered in terms of the initial density inhomogeneities present during the supernova explosion. The eastern and northern regions possibly had lower densities that could have been dispersed by the blast-wave in the initial expansion phase. On the other hand, the density of the medium towards the western and southern sides were higher. This could have led to compression and triggering of star-formation.  Alternately, it is possible that star-formation activity in the molecular cloud(s) preceded the supernova explosion due to the rapid evolution of a massive star, and the clouds were dislodged by the associated pressure of the blast wave. 

We attempt to estimate the age of the SNR based on simple analytic calculations. It is widely believed that after the initial supernova explosion, the ejecta expands out freely in the interstellar medium at supersonic speeds. This expansion continues until the SNR sweeps up its own mass. And once the swept up mass exceeds the ejecta mass, the object enters a second adiabatic phase that can be explained using the blast wave solutions proposed by Sedov \citep{1959sdmm.book.....S}. For the expansion of a blast wave, the SNR age can be estimated using the expression given by \citet[e.g.,][]{1972ApJ...178..159C}:

\begin{equation}
R \,(\textrm{pc}) = 12.9\,(\epsilon_0/n_0)^{1/5}\,t_4^{2/5}
\end{equation}

\noindent Here, $R$ is the radius of the SNR, $\epsilon_0$ is the explosion energy in units of $0.7\times10^{51}$~erg, $n_0$ is the ambient density in units of cm$^{-3}$ and $t_4$ is the age of the remnant in 10$^4$~yr. For a distance of 2~kpc, we consider the mean radius of SNR as  4~pc (Sect.3.1). For a typical average ambient density of 10~cm$^{-3}$ and assuming $\epsilon_0$=1, we calculate the age of the SNR as $2.4\times10^3$~yr. If we consider $n_e=300$~cm$^{-3}$, estimated from temperature of the warm dust component residing in plasma (Sect. 3.4.1) as the ambient density, the age estimate is larger by a factor of $\sim5$, i.e. $9.4\times10^3$~yr. Although there are uncertainties involved in the density estimates, we presume that the age of this SNR is a few thousand years. A further increase in an order of magnitude in density, i.e $10^4$~cm$^{-3}$ leads to an age estimate of  $3.8\times10^4$~yr

\par Considering the close proximity of the embedded clusters and \hii~regions, we next investigate whether the star formation in  IRAS 17256--3631 and IRAS 17258--3637 could have been triggered by the supernova or not. These \hii~regions appear  embedded in the radio emission from the SNR. If the \hii~regions are triggered by this SNR, the remnant ought to be much older than the \hii~regions. A typical lower limit to the age of the ultracompact \hii~regions based on chemical clocks is $\sim10^5$~yr \citep{2014A&A...569A..19T} and for later stages such as compact \hii~regions, it is even larger. This is an order of magnitude larger than the age estimate of this SNR. The age of young stellar objects towards IRAS~17256-3631 is found to be the order of $10^5-10^6$~yrs \citep{2016MNRAS.456.2425V}. Thus, it appears that the star-formation in the molecular cloud is unlikely to have been triggered by the supernova explosion. However, it is also possible that some star formation activity is possibly triggered \citep[e.g., Clump C4 in IRAS 17256--3631; see][]{2016MNRAS.456.2425V}. The SNR progenitor is likely to have been an older generation star in the giant molecular cloud associated with the nearby star forming complexes. In order to find the progenitor of the SNR, we have searched for possible pulsars associated with the SNR in the ATNF pulsar catalog \citep{2004IAUS..218..139H}. We could not find any pulsar that has been detected in this region. 

\subsection{Alternate explanations for SNR G351.7--1.2}

We have attributed the emission from this region to a previously unknown SNR G351.7--1.2. In this section, we explore alternate scenarios that could give rise to the observed radio, infrared  and optical emission. Apart from SNRs, emission from \hii~regions, planetary nebulae and stellar wind bubbles can also produce emission in optical to radio wavebands. A major tool to distinguish the different classes of objects is based on analysing the emission spectra. For example, the radio spectra of planetary nebulae and \hii~regions are predominantly thermal \citep[e.g.,][]{1968IAUS...34...87T}. The non-thermal radio spectral index of $-0.8$ rule out the \hii~region and planetary nebula hypotheses. In the case of a stellar wind bubble, the central star is most likely to be a Wolf-Rayet star \citep{1990MNRAS.246..358I} and non-thermal radio emission has been reported around many Wolf-Rayet stars \citep[e.g.,][]{1999IAUS..193...59C}. If the radio and optical emission is due to a Wolf-Rayet star, one would expect its central star to be observable (apparent magnitude $<$~10 at a distance of 2~kpc). The optical magnitudes of all stars that are located within 30$\arcsec$ of the H$\alpha$ shell are 14 magnitudes or fainter. Either the emission from central star is heavily obscured by the foreground ISM, or it is likely to be SNR. Based on our observational analysis and possible association with $\gamma$-ray source, we adhere to the SNR hypothesis. Follow up optical spectral line observations are envisaged to ascertain the SNR nature of this source. 

\begin{figure}
\centering
\hspace*{-0.9cm}
\includegraphics[scale=0.21]{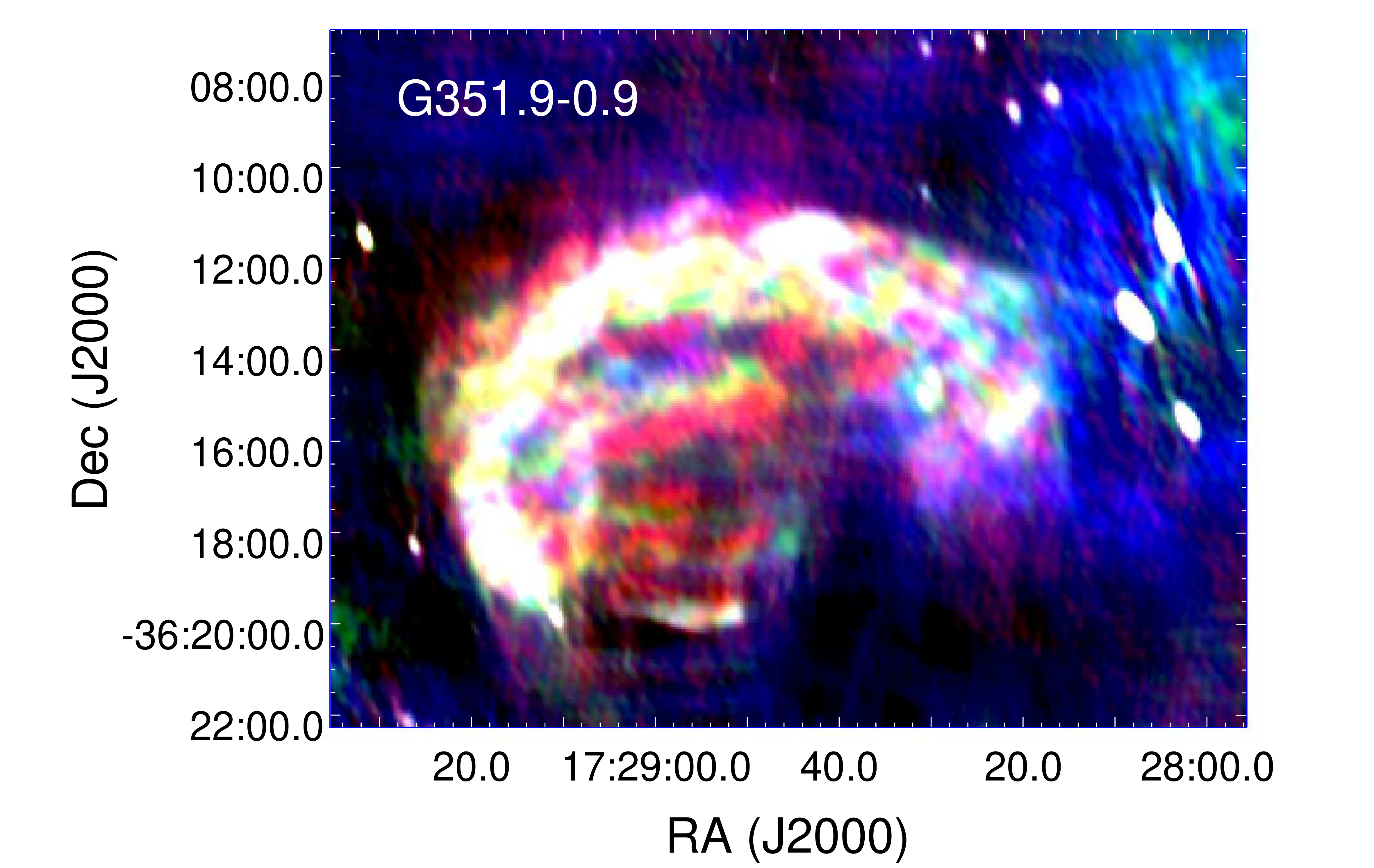}
\caption{Three color composite image of the SNR~G351.9--0.9. The 321~MHz radio map is shown in red, 385~MHz map in green and 450~MHz map in blue.}
\label{snr351p9}
\end{figure} 

\subsection{Other SNR/SNR candidates in the field of view}

Apart from SNR~G351.7, we have detected other SNR candidate(s) in the field of view. These are briefly discussed in this section. 

\par SNR~G351.9--0.9 has been previously catalogued as a Galactic SNR \citep{2014BASI...42...47G}. This SNR is located to the north of ($\sim25\arcmin$ away) SNR~G351.7. The 321~MHz image of this SNR is shown in Fig.~\ref{snr351p9}. This SNR exhibits a shell morphology consistent with the morphological classification given in the catalogue. The radio emission spans a region $13.0\arcmin\times8.4\arcmin$. A bright arc-like feature is visible in the north-east direction where the SNR exhibits a sharp boundary. The emission appears to be patchy towards the interior of the arc. The flux density of this source at 321~MHz is 10.7~Jy. From the spectral tomography map discussed in Sect. 3.2.1, we have identified steeper spectral index features ($\alpha<-0.4$) in this SNR. This is consistent with non-thermal emission expected from the SNR. The average spectral index of this SNR is found to be $-1.2\pm0.4$.

\par
\begin{figure}
\centering
\hspace*{-0.8cm}
\includegraphics[scale=0.34]{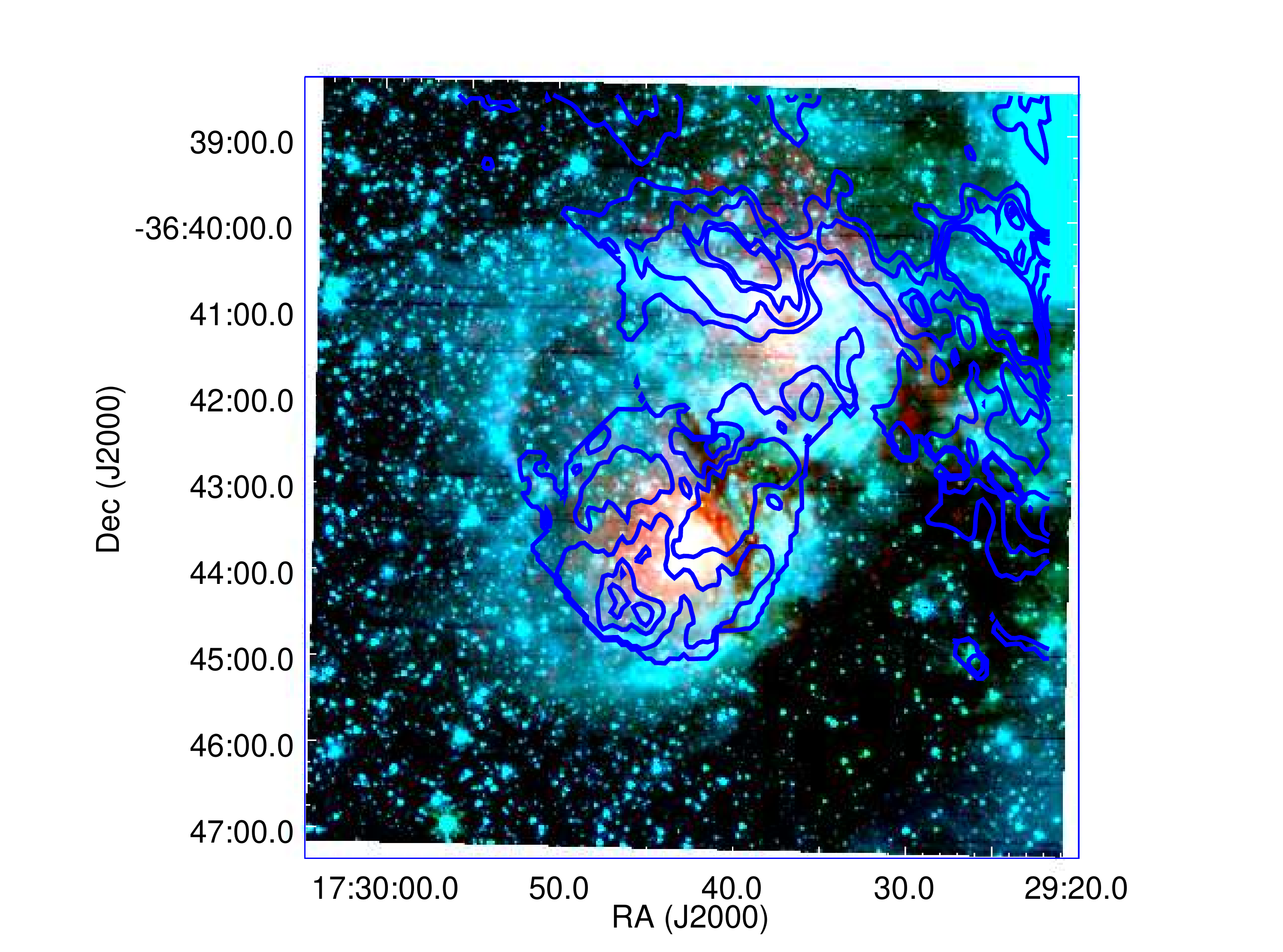}
\caption{Three color composite image of the bubble G351.6--1.3. The MSX 21.3~$\mu$m map tracing warm dust is shown in red, \textit{Spitzer IRAC} 4.5~$\mu$m and 3.6~$\mu$m maps in green and blue. Overlaid is the 321--480 steep spectral tomography contours ($\alpha<-0.4$). }
\label{bubble}
\end{figure}

 An alternate spherical structure visible in the radio bands is the source G351.6--1.3, that is located towards the south-east of the SNR~G351.7 (enclosed within aperture A6). Based on mid-infrared morphology and its association with neighbouring \hii~regions, \citet{2014ApJS..212....1A} classified this source as a group \hii~region candidate. This source is nearly spherical with an angular diameter of $1.5\arcmin$. As no distance information is available for this region ,we could not calculate the physical size of the source. The spectral index analysis of this region shows steep negative spectral index of $\alpha=-0.90$ indicating the presence of non-thermal emission. From the color composite image shown in Fig.~\ref{colcom}, we identify significant infrared emission associated with the radio source in the form of filamentary structures encompassing the radio source. We also detect infrared emission towards the centre of the object. No optical emission has been detected towards this region. An enlarged view of this source is presented in Fig.~\ref{bubble}. In the image, the MSX 21.3~$\mu$m is shown in red, \textit{Spitzer} IRAC 4.5~$\mu$m in green and 3.6~$\mu$m in blue. Here, we again see prominent features in mid-infrared towards the centre of the source surrounded by bright filamentary structures. If this object has a supernova origin similar to SNR~G351.7--1.2, the infrared emission is expected to be weak. But in this case, we detect strong mid-infrared emission in the outer rim of radio emission as well as towards the center. Similar structures are often associated with OB stars whose stellar winds and ionising pressure induce formation of shell/bubble like structures \citep{{1977ApJ...218..377W},{1980ApJ...238..860S}}. Thus, it is likely that the object is an expanding stellar bubble. We have also overlaid the spectral tomography contours on the color composite image to identify regions whose spectral index is negative, steeper than $-0.4$ (non-thermal emission). The contours show a spherical structure with an extended tail towards the north, pointing to the SNR. The peak of the tomography image is shifted southward of the radio continuum peak. The source could possibly be a bubble showing signatures of non-thermal emission similar to other stellar bubbles where previous studies have reported signatures of non-thermal emission \citep[e.g.,][]{{2016AJ....152..146N},{2017MNRAS.472.4750D}}. The non-thermal emission from such sources has been attributed to shocks induced by outflows and/or winds \citep[e.g.,][]{1996ApJ...459..193G}. The source could also be a non-thermal source in the foreground/background which is not associated with the bubble. There is also a possibility of G351.6--1.3 being a supernova remnant within a wind blown bubble as in the case of RCW86 \citep{2011ApJ...741...96W}. As there is limited information, we refrain from commenting further about the nature of this source.

\section{Conclusion}
We present low frequency radio observations of the star forming complex associated with the \hii~regions IRAS 17256--3631 and IRAS 17258--3637 using uGMRT. We have identified a new ionised shell of radius $\sim14\arcmin$ encompassing the two \hii~regions. The shell has an opening towards the north-east direction. Our radio spectral index analysis point towards the non-thermal nature of radio shell. The estimated spectral index of this object is $-0.79$. We also find evidences for spectral index variations across this shell. A morphologically similar optical counterpart is detected in the H$\alpha$ image. We believe that the radio emission and its optical counterpart are likely to be tracing the emission from a previously unidentified supernova remnant SNR~G351.7. Shock heated mid-infrared dust emission is seen towards the region and cold dust emission is only detected towards the west of the SNR candidate that could explain the partially broken-shell morphology of this object. A $\gamma$-ray source 1FGLJ1729.1--3641c is located towards the south-west of the radio shell and is likely associated with the SNR itself. An infrared bubble-like structure G351.6--1.3 is located towards south-east of SNR~G351.7. The radio emission from this bubble exhibits signatures of non-thermal emission.

\bigskip
\noindent \textbf{ACKNOWLEDGEMENT}\\

\par We thank the referee for valuable comments and suggestions which helped in improving the quality of the paper. We thank the staff of GMRT, who made the radio observations possible. GMRT is run by the National Centre for Radio Astrophysics of the Tata Institute of Fundamental Research.

\bibliography{refe}

\end{document}